\documentclass[review]{elsarticle}
\usepackage{lineno,hyperref}
\modulolinenumbers[5]

\usepackage{caption}
\usepackage{subcaption}
\usepackage{graphicx}
\usepackage{graphicx}
\usepackage{xspace}
\usepackage{color}
\usepackage{braket}
\usepackage{multirow}
\usepackage{amsmath}

\newcommand{\openatom}{\textsc{OpenAtom}}

\newcommand{\charmpp}{\textsc{Charm}{\tt++}}

\newcommand{\CC}{C\kern -0.0em\raise 0.5ex\hbox{++}\xspace}

\definecolor{MyBlue}{rgb}{0.1,0.3,0.75}

\begin{document}

\begin{frontmatter}

\title{Scalable GW software for quasiparticle properties using OpenAtom}

\author[yale]{Minjung Kim\fnref{myfootnote}}
\author[yale]{Subhasish Mandal\fnref{equal}}
\fntext[equal]{These two authors contributed equally.}
\author[UIUC]{Eric Mikida}
\author[UIUC]{Kavitha Chandrasekar}
\author[UIUC]{Eric Bohm}
\author[UIUC]{Nikhil Jain}
\author[UIUC]{Qi Li}
\author[IBM,Pim]{Glenn J. Martyna}
\author[UIUC]{Laxmikant Kale}
\author[yale]{Sohrab Ismail-Beigi\corref{corrauthor}}
\cortext[corrauthor]{Corresponding author}
\ead{sohrab.ismail-beigi@yale.edu}

\address[yale]{Department of Applied Physics, Yale University, New Haven, Connecticut 06520, USA}
\address[UIUC]{Department of Computer Science, University of Illinois at Urbana Champaign, Urbana, Illinois 61801, USA}
\address[IBM]{IBM T.J. Watson Research Center, Yorktown Heights, New York 10598, USA}
\address[Pim]{Pimpernel Science, Software and Information Technology, Westchester, NY 10598, USA}

\begin{abstract}
The GW method, which can describe accurately electronic excitations, is one of the most widely used {\it ab initio} electronic structure technique and allows the physics of both molecular and condensed phase materials to be studied.
However, the applications of the GW method to large systems require supercomputers and highly parallelized software to overcome the high computational complexity of the method scaling as $O(N^4)$. Here, we develop efficient massively-parallel GW software for the plane-wave basis set by revisiting the standard GW formulae in order to discern the optimal approaches for each phase of the GW calculation for massively parallel computation. These best numerical practices are implemented into the \openatom{} software which is written on top of \charmpp{} parallel framework. We then evaluate the performance of our new software using  range of system sizes. Our GW software shows significantly improved parallel scaling compared to publically available GW software on the Mira and Blue Waters supercomputers, two of largest most powerful platforms in the world. 

\end{abstract}

\begin{keyword}
electronic structure method, GW approximation, parallel software
\end{keyword}

\end{frontmatter}

\section{Introduction}

The ability to predict the properties of materials from first principles permits one to theoretically understand and design the novel functional molecules and materials without recourse to experiment or employing empirical techniques.  Density functional theory (DFT)~\cite{hohenberg_inhomogeneous_1964,kohn_self-consistent_1965} represents a powerful and computationally effective first principles method for computation of molecular and solid state materials properties and is the most widely used technique to compute the ground-state properties of large molecules and/or condensed phase systems. DFT provides a solid workhorse
 for modeling condensed matter, chemical, or biological systems and leads to a highly satisfactory description of total energies, bond lengths, vibrational modes, energy barriers, etc.  However, DFT is a ground-state theory that describes the lowest energy state of the electrons in a system; the Kohn-Sham band energy spectrum of DFT does not have direct physical meaning, and consequently DFT quasiparticle properties have large quantitative errors when used to predict electronic excitations~\cite{perdew_density-functional_1982, lundqvist_theory_2013, anisimov_first-principles_1997}.

One of the most accurate and fully {\it ab initio} methods for predicting electronic band structures is the GW approximation to the electron self-energy~ \cite{hedin_new_1965,HL, aryasetiawan_gw_1998,onida_electronic_2002} which is often used to correct approximate DFT results. 
The GW approximation, also called the GW method, is an {\it ab initio} quasiparticle approach that computes the effects of an important set  of electron-electron interactions and was introduced by Hedin in 1965~\cite{hedin_new_1965}; a number of reviews and summaries of GW are available~\cite{hedin_effects_1970,HL,aryasetiawan_gw_1998,onida_electronic_2002}.  

An important research topic in the GW community has been to find algorithms that reduce the computational cost of the method. 
GW approaches have been introduced that scale as $O(N^4)$ but have smaller scaling prefactors because they avoid the use of unoccupied states~\cite{wilson_efficient_2008, wilson_iterative_2009, rocca_ab_2010,lu_dielectric_2008, giustino_gw_2010,umari_gw_2010,govoni_large_2015} or because they use sum rules or energy integration to greatly reduce the number of unoccupied states~\cite{bruneval_accurate_2008,berger_ab_2010,gao_speeding_2016}. Several strategies have been presented to create cubic-scaling $O(N^3)$ methods: a spectral representation approach~\cite{foerster_on3_2011}, a space/imaginary time method~\cite{liu_cubic_2016} requiring analytical continuation from imaginary to real frequencies, and our own contribution on cubic scaling GW that works directly with real frequencies~\cite{our_cubic_scaling}.  Finally, linear scaling $O(N)$ GW is possible via stochastic sampling methods~\cite{neuhauser_breaking_2014} for the total density of electronic states within GW (the non-deterministic stochastic noise must be added to the list of usual convergence parameters).  We note that all these reduced scaling approaches derive their acceleration by working in real space (as opposed to reciprocal or Fourier space).

In this paper, a number of new and useful advances are presented.  First, an analysis of the computational advantage of using real space versus reciprocal space for the different stages of a standard GW calculation under a plane-wave basis set is provided in order to identify the optimal approach for each stage.  We show that even for a standard $O(N^4)$ GW plane-wave approach, the judicious use of $r$-space can be highly beneficial.  Second, we present a new GW software application based on our analysis for the community to use for  prediction, validation, and scientific investigation.  Third, our GW software is parallelized using  modern virtualization concepts enabled by the  \charmpp\ parallel middleware~\cite{charm_2014} which, as we demonstrate, leads to excellent and efficient parallel scaling to very large numbers of parallel processing units compared to standard MPI-based applications --- the wall clock time to solution for a fixed problem size is reduced via efficient massive parallelization of the method.

\section{Defining equations}
%%\input{equations}

% Defining equations part

The GW method, more properly the GW approximation to the electron self-energy, is an {\it ab initio} quasiparticle approach designed to include the effects of electron-electron interactions into a basic band structure method such as Density Functional Theory (DFT). 
We will now highlight the main equations of interest for our methodology.  The theoretical object of interest is the one-electron Green's function $G(x,t,x',t')$, which describes the propagation amplitude of an electron starting at $x'$ at time $t'$ to $x$ at time $t$~\cite{negele_quantum_1998}:
\[
iG(x,t,x',t') = \left\langle T\left\{\, \hat \psi (x,t) \, \hat \psi(x',t')^\dag \, \right\} \right\rangle\,,
\]
where the electron coordinate $x=(r,\sigma)$ specifies electron position ($r$) and spin ($\sigma$).
Here, $\hat \psi(x,t)$ is the electron annihilation field operator at $(x,t)$, $T$ is the time-ordering operator, and the average is over the statistical  ensemble of interest.  For this work, we focus on the zero-temperature case (i.e., ground-state averaging).  Knowledge of the Green's function permits computation of all one-particle operator averages as well as quasiparticle energies, wave functions, and spectral properties.  For materials described by a time-invariant Hamiltonian, the Green's function in the frequency domain obeys Dyson's equation
\[
G^{-1}(\omega) = \omega I  - \left[ T + V_{ion} + V_H + \Sigma(\omega) \right]
\]   
where we have suppressed the $x,x'$ indices to write the equation compactly in matrix form.  Above, $I$ is the identity operators, $T$ the electron kinetic operator, $V_{ion}$ the electron-ion interaction potential operator (or pseudopotential for valence electron only calculations), $V_H$ is the Hartree potential operator, and $\Sigma(\omega)$ is the self-energy operator encoding all the many-body interaction effects on the electron Green's function.  

The GW approximation~\cite{hedin_new_1965} is a specific approximation for the self-energy given by 
\[
\Sigma(x,x',t) = iG(x,x',t)W(r,r',t^+)
\] 
where $t^+$ is infinitesimally larger than $t$ and $W(r,r',t)$ is the dynamical screened Coulomb interaction between an external test charge at $(r',0)$ and $(r,t)$:
\[
W(r,r',\omega) = \int dr''\ \epsilon^{-1}(r,r'',\omega)V_c(r'',r')\,.
\]
Here, $\epsilon(r,r',t)$ is the linear response dynamic and nonlocal microscopic dielectric screening matrix and $V_c(r,r')=1/|r-r'|$ is the bare Coulomb interaction.  As such, the GW self-energy includes the effects due to dynamical and nonlocal screening on the propagation of electrons in a many-body environment.  

To provide a complete set of equations, one must approximate $\epsilon$, and the most common approach is the random-phase approximation (RPA):  one first writes $\epsilon$ in terms of the irreducible polarizability $P$ via
\[
\epsilon(r,r',\omega) = \delta(r-r') - \int dr''\  V_c(r,r'') P(r'',r',\omega)
\]
and then relates $P$ back to the Green's function by the RPA
\[
P(r,r',t) = \sum_{\sigma,\sigma'} -iG(x,x',t)G(x',x,-t)\,.
\]

In the vast majority of GW calculations, the Green's function is approximated by an independent electron (band theory) form specified by a complete set of one-particle eigenstates $\psi_n(x)$ and eigenvalues $E_n$
\begin{equation}
G(x,x',\omega) = \sum_n \frac{\psi_n(x)\psi_n(x')^*}{\omega - E_n}\,.
\label{eq:Gindep}
\end{equation}
The $\psi_n$ and $E_n$ are typically obtained as eigenstates of a non-interacting one-particle Hamiltonian from a first principles method such as DFT although one is not limited to this choice.

For our purposes, the frequency domain representations of all quantities are most useful.  The Green's function $G$ is already written above in Eq.~(\ref{eq:Gindep}).  The polarizability $P$ is
\[
P(r,r',\omega) = \sum_{c,v,\sigma,\sigma'} \frac{2(E_c-E_v)\psi_c(x)\psi_v(x)^*\psi_c(x')^*\psi_v(x')}{\omega^2-(E_c-E_v)^2}
\]
where $v$ labels occupied (valence) eigenstates while $c$ labels unoccupied (conduction) eigenstates.  We will be specifically interested in the static ($\omega=0$) polarizability
\begin{equation}
P(r,r') = 2 \sum_{c,v,\sigma,\sigma'} \frac{\psi_c(x)\psi_v(x)^*\psi_c(x')^*\psi_v(x')}{E_v-E_c}\,.
\label{eq:Pstatic}
\end{equation}
Formally, the screened interaction $W$ can always be represented as a sum of ``plasmon'' screening modes indexed by $p$:
\begin{equation}
W(r,r',\omega) = V_c(r,r') + \sum_p  \frac{ 2\omega_p\, B_p(r,r') } { \omega^2 - \omega_p^2}
\label{eq:Wplasmonrep}
\end{equation}
where $B_p$ is the mode strength for screening mode $p$ and $\omega_p>0$ is its frequency.  This general form is relevant when making computationally efficient plasmon-pole models~\cite{hedin_effects_1970}.  The self-energy is then given by
\begin{equation}
\begin{aligned}
\Sigma(x,x',\omega) = -\sum_v \psi_v(x)\psi_v(x')^* W(r,r',\omega-E_v) \\ + \sum_n \psi_n(x)\psi_n(x')^*\sum_p \frac{B_p(r,r')}{\omega-E_n-\omega_p}\,.
\label{eq:Sigmafreq}
\end{aligned}
\end{equation}

Due to the complex form of the dynamic (i.e., $\omega$-dependent) self-energy which leads to high computational loads, there have been a number of approaches for simplifying the GW method that still produce reliable results. One of the successful approaches is the ``COHSEX'' approximation~\cite{hedin_new_1965} which simplifies equation (\ref{eq:Sigmafreq}) by neglecting frequency dependence and using the $\omega\rightarrow 0$ limit: 
\begin{equation}
\begin{aligned}
    \Sigma(x,x') = -\sum_v \psi_v(x)\psi_v(x')^*W(r,r',\omega=0) \\ +  \frac{1}{2}\delta(x-x')[W(r,r',\omega=0)-v(r,r')].
\label{eq:COHSEX}
\end{aligned}
\end{equation}
In addition to being quite satisfactory when used self-consistently~\cite{bruneval_effect_2006}, the accuracy of COHSEX can be improved significantly with minor adjustments while keeping its static format~\cite{kang_hybertsen_2010}.  In addition, COHSEX provides an excellent starting point for performing dynamic self-energy calculations~\cite{bruneval_effect_2006}.  Hence, the COHSEX self-energy will play a prominent part in our work below.

The core equations (\ref{eq:Gindep},\ref{eq:Pstatic},\ref{eq:Wplasmonrep},\ref{eq:Sigmafreq},\ref{eq:COHSEX}) are computationally intensive for large systems which can be ameliorated by using massively parallel computing.  The technical issues to be addressed are (1) the truncation of the band summations ($c$ in Eq.~(\ref{eq:Pstatic}) and $n$ in Eq.~(\ref{eq:Sigmafreq})), (2) whether the real space ($r$ or $x$) is the optimal representation in which evaluate parts of the calculations, (3) the potentially large sizes of the matrices involved (e.g., $P$ or $\epsilon$) as well as the large inversion problem when going from $\epsilon$ to $\epsilon^{-1}$, (4) and the computation and effective truncation of the sum over screening modes $p$ in Eq.~(\ref{eq:Sigmafreq}).

\section{Overview}

% Adding logic flow here
\begin{figure}
    \centering
    \includegraphics[width=0.5\textwidth]{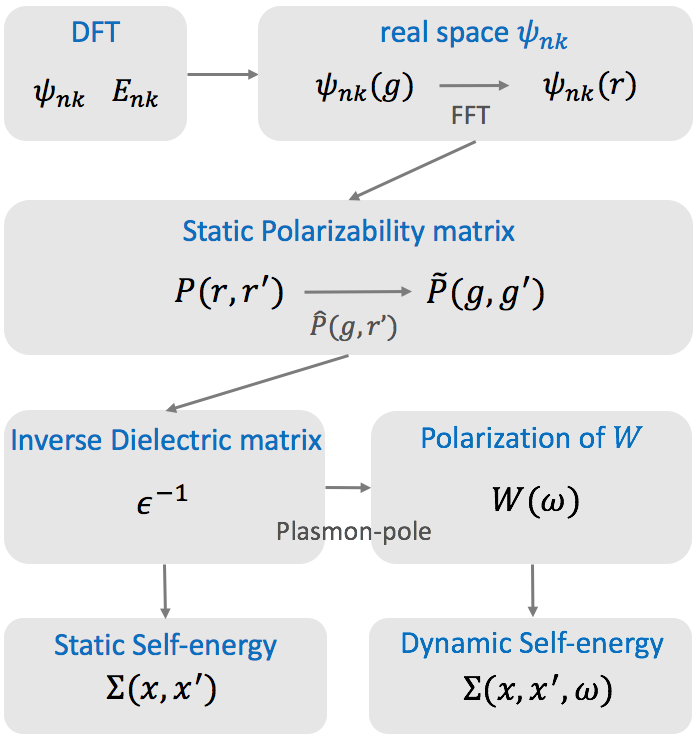}
    \caption{Workflow of GW calculation in the \openatom\ package.}
    \label{fig:flowchart}
\end{figure}

Figure~\ref{fig:flowchart} illustrates the major computational elements for the GW approach as performed by our implementation in \openatom{}. Prior to describing and analyzing the computational complexity of the various elements in detail, we provide a short overview of these elements to help guide the reader through the remainder of this paper.

\begin{enumerate}

    \item DFT: Prior to performing a GW calculation, one first carries out DFT calculations to obtain input single-particle (Kohn-Sham) electronic wavefunctions ($\psi_{nk}$) and energies ($E_{nk}$). For \openatom{} GW, these can be obtained by either ground-state \openatom{} calculations or via  calculations with the widely used Quantum Espresso software package~\cite{QE} by using a simple converter utility software we have created to change the output data format to be compatible with \openatom. 
    
    \item Static polarizability: The first step of a GW calculation is to form the static (zero frequency) polarizability matrix $P$ of Eq.~\ref{eq:Pstatic}. We calculate $P$ in $r$-space and then perform fast Fourier transforms (FFTs) to obtain $P$ in $g$-space. Detailed justification of the choice of calculating $P$ matrix in this particular way is found in Sec.~\ref{sec:staticP}. Since the $P$ calculation is the main bottleneck of all GW calculations, we also describe in detail  the parallelization strategy for its calculation in Sec.~\ref{sec:scaling}.
    
    \item Dielectric matrix: Once the $P$ calculation is completed, the static dielectric screening matrix is computed and then is inverted by iterative matrix inversion method as discussed in Sec.~\ref{sec:staticeps}. For the dynamic part of the screened interaction, we bypass an exact evaluation of the frequency-dependent $\epsilon(\omega)$ (to avoid many costly $P(\omega)$ calculations over the dense grid of $\omega$) by the plasmon-pole approximation described in  Sec.~\ref{sec:pp}.
    
    \item Self-energy:  Starting from the static inverse dielectric matrix, we can choose two separate ways to obtain the electron self-energy: static self-energy via the COHSEX approximation or the more accurate dynamic self-energy. Within the COHSEX approximation, the static screened exchange and static Coulomb hole terms are evaluated using the static inverse dielectric matrix. Unlike the static $P$ calculation, here we employ a $g$-space based method for the static self-energy calculations: this choice is discussed in Sec.~\ref{sec:staticsigma}. The calculation of dynamic self-energy via the plasmon-pole method is described in Sec~\ref{sec:dynamicsigma}.

\end{enumerate}

\section{Static polarizability}\label{sec:staticP}

The most time consuming part of a  GW calculations is the computation of the polarizability matrix of Eq.~(\ref{eq:Pstatic}). Our approach is to compute $P$ in real space, as opposed to the traditional method of computing it in reciprocal space.  In Sec.~\ref{sec:scaling}, its  implementation in \openatom\ using the \charmpp\ parallel middleware is described.

\subsection{Real space versus reciprocal space: scaling and trade offs}
% Real space
In a system with periodicity, the one-particle states and their eigenvalues have an additional quantum number (label) of $k$, the Bloch momentum.  The Bloch state $\psi_{nk}(x)$ is related to its periodic part $u_{nk}(x)$ via the standard relation
\[
\psi_{nk}(x) = \frac{e^{ik\cdot r}}{\sqrt{N_k}}u_{nk}(x)
\]
where $N_k$ is the number of $k$-points used to sample the first Brillouin zone.
The real space representation of the static polarizability matrix is given by
\begin{equation}
P^q(r,r')= -\frac{2}{N_k}\sum_{v,c,k,\sigma,\sigma'}\frac{ u_{ck}(x)^* u_{vk+q}(x) u_{vk+q}(x')^* u_{ck}(x')}{E_{ck}-E_{vk+q}}
\label{eq:staticpol}
\end{equation}
where $q$ labels the the Bloch wave vector (momentum) transfer and the entire full polarizability is given by
\[
P(r,r') = \frac{1}{N_k}\sum_q P^q(r,r')e^{iq\cdot(r-r')}\,.
\]
To calculate $P^q$ in real space, one needs real space wave functions from the mean-field calculations (typically from DFT).   \openatom\ utilizes a plane-wave basis to describe the wave functions in reciprocal space $\tilde u_{nk}(g,\sigma)$ where $g$ labels a reciprocal lattice vector, $\sigma$ the spin index, and a finite basis set is defined by a spherical cutoff condition $|g|<g_{max}^\psi$.  Thus, the first step is to transform the wave functions to real space using fast fourier transforms (FFTs).  We denote this operation via  
\[
u_{nk}(x) = FFT\left[\tilde u_{nk}(g,\sigma)\right] = \sum_g \tilde u_{nk}(g,\sigma)\, \frac{e^{ig\cdot r}}{\sqrt{\Omega}}
\]
where $\Omega$ is volume of the simulation cell.  (We have suppressed the spin index $\sigma$ for clarity.)  The inverse  to FFT is denoted as IFFT below.
We store all $u_{n,k}(r)$ for all $n$ and $k$, in memory in a distributed fashion and discard the $\tilde u_{n,k}(g)$ which are no longer needed.

Next, we form vectors in $r$ with four indices $k$, $q$, $c$, and $v$ by a point-wise multiplication of a pair of wave functions,
\begin{equation}
f_{kqcv}(x)=u^*_{vk+q}(x)u_{ck}(x)\sqrt{\frac{2}{E_{ck}-E_{vk+q}}}.
\label{eq:fdef}
\end{equation}
The real space $P^q(r,r')$ matrix is then computed via a large outer product using the $f_{kqcv}$:
\begin{equation}
P^q(r,r') = -\frac{1}{N_k}\sum_{v,c,k,\sigma,\sigma'}f_{kqcv}(x)^*f_{kqcv}(x')\,.
\end{equation}
Once the formation of $P^q(r,r')$ is complete, we then transform $P^q$ from $r$-space to $g$-space for subsequent steps involving the Coulomb interaction which is diagonal in $g$-space.  This means we must FFT both the rows and columns of $P^q(r,r')$.  The operation has two phases:  first, the FFTs are applied to the rows of $P^q(r,r')$ to generate an intermediate $\hat P^q(g,r')$ matrix
\[
\hat P^q(g,r') = IFFT_r\big[P^q(\{r\},r')\big]  \qquad \mbox{(column-wise FFT)}
\]
and then FFTs are applied to the columns of $\hat P^q$ to generate the final $g$-space polarizability
\[
\tilde P^q(g,g') = IFFT_{r'}\big[\hat P^q(g,r')\big]  \qquad \mbox{(row-wise FFT)}\,.
\] 

% g-space
In comparison, the more traditional method is to compute $P$ directly in $g$-space via the  Alder and Wiser formulae~\cite{Alder62,Wiser63},
\begin{equation}
\tilde P^q(g,g') = \frac{2}{N_k} \sum_{v,c,k,\sigma,\sigma'}\frac{M_{kqcv}(g)  M_{kqcv}(g')^* }{E_{vk+q}-E_{ck}}\,,
\label{eq:gpsaceP}
\end{equation}
where the matrix element $M_{kqcv}(g)$ are given by the Fourier transform of the product of two wave functions,
\[
M_{kqcv}(g)  = IFFT\big[u^*_{ck}(r)u_{vk+q}(r)\big].
\]
(Again, spin indices have been suppressed.)  
The traditional method works directly in $g$-space and deliver $\tilde P^q$ in that space, but requires a quadratic number of FFTs to compute the matrix element $M_{kqcv}$.

\begin{table}
\begin{center}
\begin{tabular}{| c | c | c |}
\hline
{Approach} & {Task} & {Operation Count} \\
\hline

\multirow{4}{*}{$r$-space} & Compute $u_{nk}$ & $(N_c+N_v)\cdot 100 N_r\ln N_r$ \\
& Compute $f_{kqcv}$ & $ N_c N_v\cdot N_r $  \\ 
& Compute $P^q$ from $f$ & $N_cN_vN_r^2$ \\
& $P^q \rightarrow \tilde P^q$ & $2N_r\cdot 100 N_r \ln N_r  $ \\
\hline
\multirow{2}{*}{$g$-space} & Compute $M_{kqcv}$ &  $N_cN_v\cdot 100 N_r\ln N_r$ \\
& Compute $\tilde P^q$ from $M$ & $N_cN_vN_g^2$\\
\hline
\end{tabular}
\caption{Operation counts for computing the  polarizability matrix $\tilde P^q_{g,g'}$ using the real space ($r$-space) and reciprocal ($g$-space) methods. For simplicity, the table shows operation counts for a single $k$-point.  We assume that a complex FFT of size $N_r$ costs $\approx 7N_r\ln N_r$ operations~\cite{NumRecip}.}
\label{tab:ncomp}
\end{center}
\end{table}

We compare the scaling the operation counts for the real and reciprocal space calculations of $\tilde P$ in 
Table~\ref{tab:ncomp}.  In the Table, $N_r$ and $N_g$ are the number of real space grid points and the number of reciprocal grid points for describing $P^q$ and $\tilde P^q$, respectively.  $N_c$ and $N_v$ are the number of unoccupied and occupied states.  We take a complex-valued FFT to cost $\sim 100 N_r\ln N_r$ operations~\cite{NumRecip}. In terms of FFTs, the real-space method is advantageous: the number of FFTs required for the real space method scales only linearly with the system size since $N_v$, $N_c$ and $N_r$ are grow linearly with the number of atoms, whereas the $g$-space has a quadratic ($N_cN_v$) number of FFTs to perform.  For large systems, this is a major computational savings.  

In the limit of large systems, both methods have a quartic scaling due to the matrix algebra required to form $P^q$ or $\tilde P^q$ via outer products of $f$ or $M$.  For this part of the computation, if we insist that both method result in a $\tilde P^q(g,g')$ having the same $g$-space grid of size $N_g$ as chosen by a standard plane wave method, then the real-space method will be more costly than the $g$-space method by a constant factor of about 4.   The reason for this difference is in that set of $G$ describing $\tilde P^q$ are typically chosen by a spherical cutoff condition $|g|<g^P_{max}$ while an FFT grid is a uniform grid on a parallelepiped.  Assuming a cubic grid, the volume of a sphere inscribed in a cube is 1.9 times smaller so that we expect $N_r \approx 1.9 N_g$ so that $N_r^2 \approx 3.6N_g^2$.  

Given wave functions $\tilde u_{n,k}(g)$ that are non-zero for reciprocal lattice vectors $g$ where $|g|<g^\psi_{max}$, $\tilde P^q(g,g')$ can be computed exactly using FFT grid-based methods for a Fourier grid corresponding to $g^P_{max}=2g^\psi_{max}$.  However, in practice, this leads to large FFT grids that are  unnecessary for physically converged results: the FFT grid size can be quite modest while still giving well converged results.  Conversely, for a fixed $g^P_{max}$, accurate results can be computed on an FFT grid whose equivalent cutoff is much smaller than $2g^P_{max}$.  We now turn the issue of FFT grid size.

%
%\begin{figure}
%\centering
%\includegraphics[width=0.2\textwidth]{fftgrid}
%\caption{Schematic of FFT grid when epsilon energy cutoff is the same as wavefunction energy cutoff. Blue and red circles indicate G space for occupied and unoccupied wavefunctions respectively when calculating $P_{G_1,G_2}$ ($G_1=G_{max}$ and/or $G_2=G_{max})$.}
%\label{fig:fftgrid}
%\end{figure}

\subsection{Size of r-space grid}
% some introductory sentence
When we form the matrix $P^q(r,r')$ of  Eq.~(\ref{eq:staticpol}), two wave functions are multiplied  at each $r$ point which leads to a more oscillatory function than each wave function.  In standard DFT calculations, describing such product accurately, e.g., for the electron density, requires a dense FFT.  However, GW calculations require far less stringent convergence criteria as we show here and thus permit much sparser FFT grids.

We have tested the convergence of the dielectric screening matrix versus the size of the FFT grids for two different physical systems: the small gap covalent bulk semiconductor Si and the large gap ionic insulator bulk MgO.  Norm conserving pseudopotentials~\cite{TM} are employed to generate the DFT ground-state density and electronic states using  the local density approximation (LDA)~\cite{CA,PZ}.  For both materials, we use the primitive 2-atom unit cell sampled with a uniform 8 k-point mesh.
For Si, the wave function cutoff $g_{max}^\psi$ corresponds to 25 Ryd while for MgO it is 50 Ryd.   For Si, this yields an electron density FFT grid that is $24\times24\times24$ while for MgO we have a $25\times 25\times 25$ gird (i.e., these grids correspond to $g^P_{max}=2g^\psi_{max}$).

\begin{figure}
\centering
\includegraphics[width=0.7\textwidth]{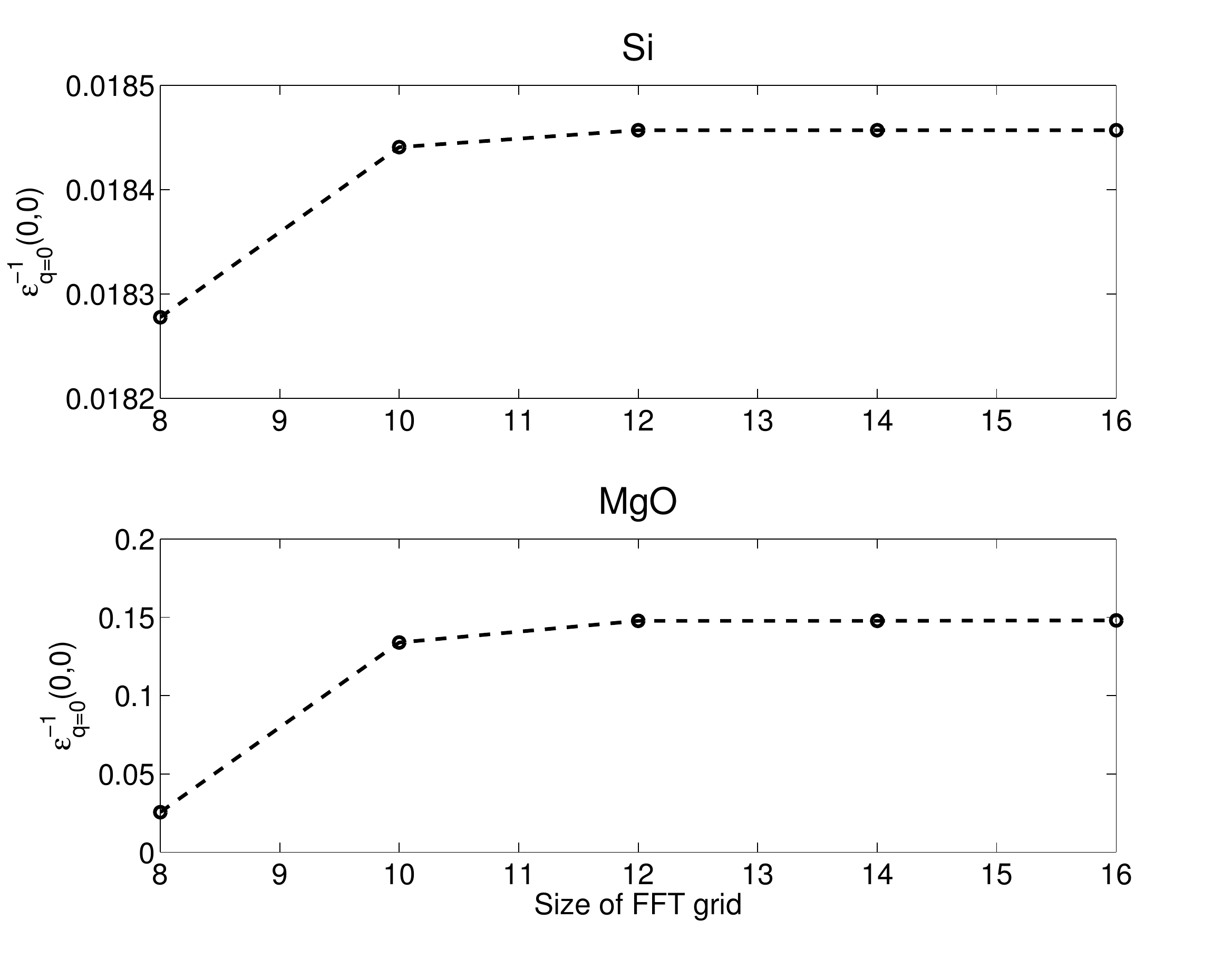}
\caption{Convergence of the macroscopic optical dielectric constant $\epsilon_\infty = 1/(\epsilon^{q=0})^{-1}_{0,0}$ with respect to the size of FFT grid used for the computation of $P^q_{r,r'}$ in real-space. For both Si and MgO, we employed 2-atom cell, 8 k points, 4 occupied and 48 unoccupied band. }
\label{fig:fftsize}
\end{figure}

Figure~\ref{fig:fftsize} shows the head of the inverse dielectric matrix at $q=g=g'=0$, $(\epsilon^{q=0})^{-1}_{0,0}$, (see the next section for its  formula) as a function of the size of the FFT  grid used for computing the $P^q(r,r')$ matrix in our real-space approach.
We observe that $(\epsilon^{q=0})^{-1}_{0,0}$ is already very well converged for a $12\times12\times12$ grid which is half the spacing of the FFT density grid in both cases.  Other matrix elements for $\epsilon^{-1}$ are also converged at least the same level for a $12\times12\times12$ grid.  These two examples show that the FFT grid for computing $P^q_{r,r'}$ can be about half as dense as the FFT density grid.  Given that $r$ is a three dimensional vector and $P^q$ is a matrix, this reduces storage and computation requirements by a factor of 64 compared to the stringent use of the full  density FFT grid.

\section{Dielectric matrix and its frequency dependence }
%%\input{epsiloninv}

% How we build epsilon and then do the inverse.  Probably nothing really deep here.
% But our plasmon pole may have big advantage when we cut off many of the high energy modes.

\subsection{Static inverse dielectric matrix}\label{sec:staticeps}

Once the static polarizability matrix is calculated, we build a symmetric static dielectric matrix
\begin{equation}
\epsilon^q_{g,g'} = \delta_{g,g'} - \sqrt{V_c(q+g)}\cdot \tilde P^q_{g,g'}\cdot \sqrt{V_c(q+g')}
\end{equation}
where $\delta_{x,y}$ is  Kronecker's delta and $V_c$ is the bare Coulomb potential in reciprocal space.  For the standard Coulomb interaction this takes the form
\begin{equation}
V_c(q+g) = \frac{4\pi e^2}{\Omega|q+g|^2}\,,
\end{equation}
where $\Omega$ is a volume of the simulation cell.
However, we keep the formalism general below and retain $V_c(q+g)$ throughout: truncated Coulomb interactions~\cite{onida_sodiumtetramer_1995,spataru_quasiparticle_2004,ismail-beigi_truncation_2006,rozzi_exact_2006} for simulations of systems with reduced periodicity  simply correspond to using a different formula for $V_c(q+g)$.  
Computing $\epsilon^q$ from $\tilde P^q$ is simple and is done in-place.

\begin{figure}
\centering
\includegraphics[width=0.7\textwidth]{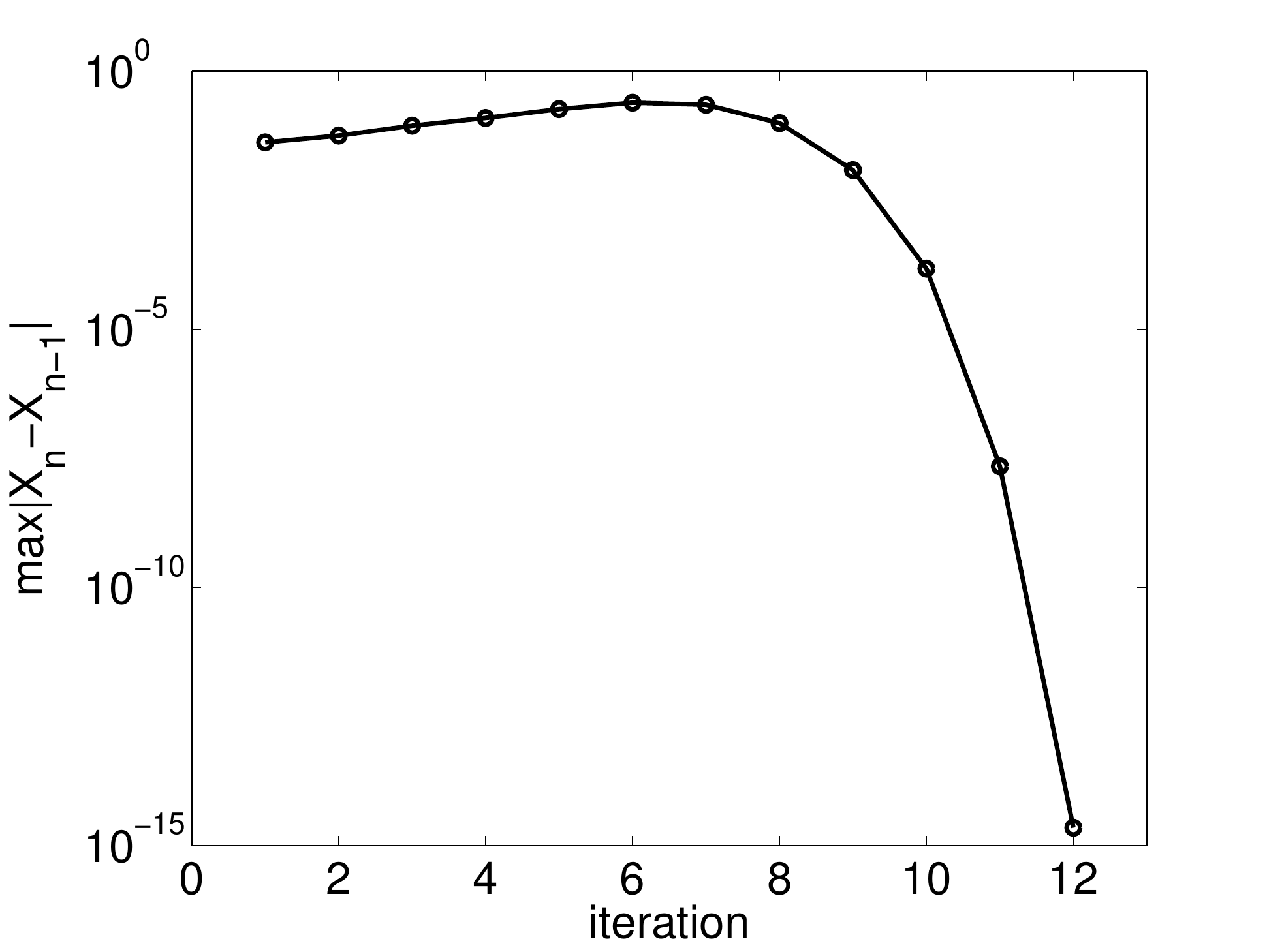}
\caption{Iterative matrix inversion method performance data for bulk Si with 2-atom unit cell with 8 k points at $q=\pi/a(1,1,-1)$. The epsilon matrix was calculated using total 52 states. Data show the largest entry of the matrix $|\textbf{X}_n - \textbf{X}_{n-1}|$ versus  iteration number.}
\label{fig:iter}
\end{figure}

The more difficult calculation is the matrix inversion of $\epsilon^q$ to $(\epsilon^q)^{-1}$ required to compute other screening properties.  We perform the inversion via a widely used iterative matrix inversion technique that relies only on Newton's method and matrix multiplication~\cite{ABI1,ABI2}.  In brief, for an arbitrary matrix $\textbf{A}$, $\textbf{A}^{-1}$ is obtained by following iterations:
\begin{equation}
\textbf{X}_{n+1} = \textbf{X}_n(2\textbf{I}-\textbf{AX}_n)\\
\end{equation}
\[
\textbf{X}_0=\alpha\textbf{A}^T, \alpha=\frac{1}{{\rm max}_{i} \sum_j(\textbf{AA}^T)_{i,j}} 
\]
\noindent This iteration is terminated when $|\textbf{X}_{n+1,(i,j)}-\textbf{X}_{n,(i,j)}|$ is below a tolerance for all $i,j$ pairs, and the last $\textbf{X}$ matrix is the approximation to $\textbf{A}^{-1}$. Figure~\ref{fig:iter} shows  numerical results for the inversion of the dielectric matrix of Si at $q=\frac{\pi}{a}(1,1,-1)$ for 2-atom system.  The cutoff for the $\tilde P^q$ creates a dielectric matrix of size $410\times410$, and the tolerance was set to $10^{-11}$.
After 12 iterations, the desired accuracy was achieved.  We note that while canonical inversion methods based on, e.g., the $LU$ decomposition, have a lower operation count than repeated full matrix-matrix multiplication, they are much harder to parallelize compared to matrix-matrix multiplication.

%---------------------------------------------------------------
\subsection{Dynamic inverse dielectric matrix}
\label{sec:pp}

Obtaining the static screening matrix $(\epsilon^q)^{-1}$ is a good start to describing electronic screening but it is not sufficient.  In general, and in particular within the GW approximation, screening is dynamical so that one needs to describe a frequency dependent inverse dielectric function $(\epsilon^q(\omega))^{-1}$.  In principle, one can generalize and repeat the above considerations to compute a dynamic $\tilde P^q(\omega)$ matrix which leads to $\epsilon^q(\omega)$ and upon inversion to $(\epsilon^q(\omega))^{-1}$.  This direct approach is computationally prohibitive as many computations of $\tilde P^q$ are required over a dense grid of $\omega$.  An alternative approach is to choose an approximate analytical form for $(\epsilon^q(\omega))^{-1}$ with adjustable parameters:  the parameters are chosen to match computed results and known physical conditions, and the analytical form avoids the need for $\omega$ sampling.

Plasmon-pole (PP) type approximations achieve these ends.  Of the many proposed such approximations, three are widely used:  that of Hybertsen and Louie (HL)~\cite{HL}, von der Linden and Horsch (vdLH)~\cite{vdLH}, and Engel and Farid~\cite{EFPP}. In \openatom, we have chosen to follow the vdLH to create a PP model: however, in our implementation we diagonalize the polarization part of the static screened interaction $W$ while vdLH diagonalize the static $(\epsilon^q)^{-1}$; otherwise, the sum rules imposed and the number of screening modes are identical.  Our choice of diagonalizing polarization part of $W$ is practical: this is the matrix that appears in the dynamic GW self-energy so diagonalizing it leads to simpler relations.

Specifically, we form the static screened interaction $W^q$ via
\[
W_{g,g'}^q = \sqrt{V_c(q+g)}\cdot  (\epsilon^q)^{-1}_{g,g'}\cdot \sqrt{V_c(q+g')}
\]
and the polarization part $S^q$ is separated off as
\[
S_{g,g'}^q = W_{g,g'}^q - V_c(q+g)\delta_{g,g'} = \sqrt{V_c(q+g)}[  (\epsilon^q)^{-1}_{g,g'}-\delta_{g,g'}]\sqrt{V_c(q+g')}\,.
\]
The Hermitian matrix $S^q$ is then written in its diagonal basis,
\begin{equation}
S^q_{g,g'} =  \sum_{\alpha} A^q_{g,\alpha}\cdot \sigma^q_\alpha\cdot  {A^q_{g',\alpha}}^*\,,
\label{eq:eigdecomp}
\end{equation}
where the eigenvalues of $S^q$ are $\sigma^q_\alpha$ and the orthonormal eigenvectors are the columns of $A^q$.  The dynamical behavior is then approximated by the PP form
\begin{equation}
S^q(\omega)_{g,g'} =  \sum_{\alpha} A^q_{g,\alpha}\cdot \frac{\sigma^q_\alpha \cdot {\omega^q_{\alpha}}^2}{{\omega^q_{\alpha}}^2-\omega^2} \cdot {A^q_{g',\alpha}}^*
\label{eq:ppform}
\end{equation}
Here, $\omega_\alpha^q$ is the frequency of the PP mode $\alpha$ frequency: this free parameter determined by applying the Johnson sum-rule~\cite{Johnson_1974,HL} which describes the integral of the screening function.  In our nomenclature, it translates into
\[
\int_0^\infty d\omega\cdot\omega\cdot\mbox{Im} S^q(\omega)_{g,g'} = -\frac{\pi}{2}\cdot
V_c(q+g)\cdot V_c(q+g')\cdot[(q+g)\cdot(q+g')]\cdot\rho_{g-g'}
\]
where $\rho_g$ is the ground-state electron density represented in $g$-space.  This means that  $\omega^q_\alpha$  is given by
\begin{equation}
{\omega_\alpha^q}^2 = \frac{1}{\sigma^q_\alpha}\sum_{g,g'} {A^q_{g,\alpha}}^*  \cdot V_c(q+g)V_c(q+g')[(q+g)\cdot(q+g')]\rho_{g-g'}
 \cdot A_{g',\alpha}^q\,.
\label{eq:ppfreq}
\end{equation}

\begin{figure}
\centering
\includegraphics[width=0.7\textwidth]{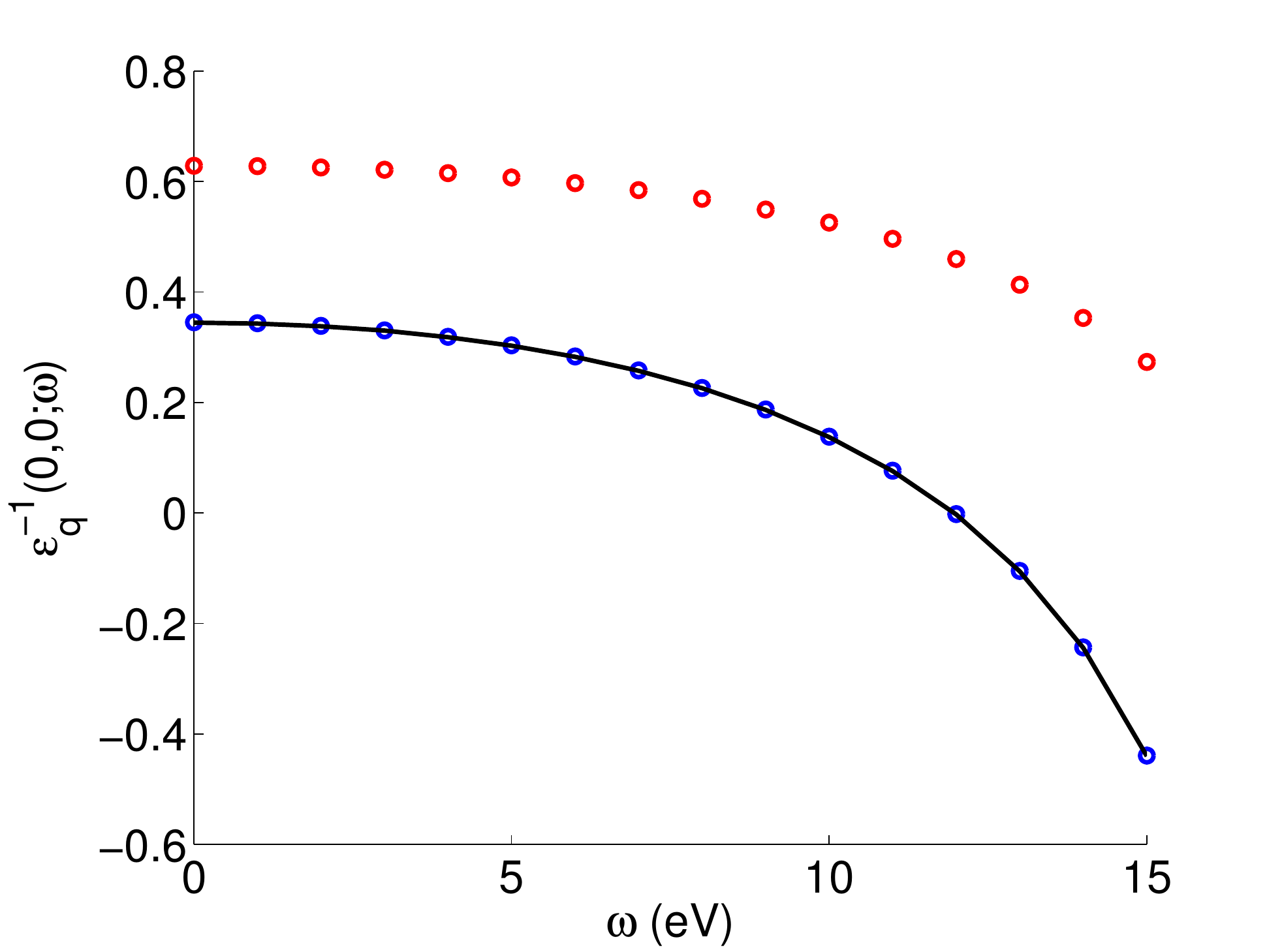}
\caption{Convergence of our plasmon-pole model $(\epsilon^q(\omega))^{-1}_{0,0}$  at $q=\frac{\pi}{a}(1,1,-1)$ for bulk Si with 2-atom unit cell with 8 k points. Red and blue dots indicate $(\epsilon^q(\omega))^{-1}_{0,0}$ constructed with one and five eigenmodes, respectively. Black line includes the result of including all eigenmodes. The size of $\epsilon^{-1}$ matrix is $410\times410$.}
\label{fig:PP}
\end{figure}

Our motivation for using the vdLH PP form compared to the HL PP form is two fold:  first, the vdLH has only $N_g$ plasmon modes compared to $N_g^2$ for the HL approach so that less computations are needed overall.  Second, the vdLH eigenrepresentation presents rapid convergence.  Figure~\ref{fig:PP} shows the convergence of the vdLH $(\epsilon^q)^{-1}_{g,g'}(\omega)$ matrix with respect to the number of eigenmodes $\alpha$ included in the sum for  bulk Si with a 20 Ryd cutoff for $\tilde P^q$ which yields $N_g=410$.  The plot shows that the dynamic dielectric response at physically important frequencies converges with a very small number of eigenmodes: with only 5 out of 410 eigenmodes,  the difference between the truncated and exact computation is already essentially invisible.  What this means is that we can use a very small number of screening modes to represent the dielectric screening thereby reducing computational efforts significantly.  And we can use efficient iterative diagonalization methods to find the few most important eigenvectors and eigenvalues of $S^q$ that feed into this procedure.

\vspace{0.2in}

\section{Static and dynamic self-energy}\label{sec:sigma}
%%\input{sigma}

%We have static and dynamic parts of the self-energy.  Write them separately and explain why a different strategy is needed to compute them compared to the P(r,r") method: explain scaling difference and some numerics.

%Hopefully, we can have a static part and a dynamic part explaining how we (plan) to do it.

Here we present our strategy for computing the GW self-energy $\Sigma(x,x',\omega)$ of Eq.~(\ref{eq:Sigmafreq}).  A significant part of the self-energy has a static behavior (i.e., has no frequency $\omega$ dependence) and is known as the ``COHSEX'' approximation~\cite{hedin_new_1965}.  The remainder of the self-energy is dynamic.  This separation allows us to first develop and refine our approach in some detail within the simpler COHSEX framework and then to apply the best computational approach for each part of the self-energy.

\subsection{Static self-energy}\label{sec:staticsigma}
%%\input{staticsigma}

 % This is the static sigma (COHSEX) section

The COHSEX approximation to the self-energy leads to a self-energy that is the sum of three separate terms with distinct physical meaning:
\begin{equation}
\Sigma_{cohsex}(x,x') = \Sigma_x(x,x') + \Sigma_{sex}(x,x') + \Sigma_{coh}(x,x')
\end{equation}
where the first term is the bare exchange (a.k.a., Fock exchange) self-energy
\[
\Sigma_x(x,x') = - \sum_{v,k} \psi_{vk}(x)\psi_{vk}(x')^*V_c(r,r')
\]
that describes the importance of fermionic behavior of electrons whereby exchanging them leads to a negative sign in the overall wave function (the minus sign in front).  The static screened exchange term
\[
\Sigma_{sex}(x,x') = -\sum_{v,k} \psi_{vk}(x)\psi_{vk}(x')^* \left [ W(r,r',\omega=0)-V_c(r,r') \right]
\]
describes the fact that the usual Fock exchange process must be screened in a solid since the bare Coulomb interaction $V_c$ should be replaced by the screened interaction $W$; the static nature is highlighted by the fact that the screened interaction is evaluated at zero frequency.  The last term
\[
\Sigma_{coh}(x,x') = \frac{1}{2}\left[W(r,r',\omega=0)-V_c(r,r')\right]\delta(r-r')
\]
is the static Coulomb hole term describing the fact that the presence of an electron at $x$ creates a depletion of the density of the other electrons (due to electrical repulsion) which creates a potential ``hole'' at the position of the electron (hence $r=r'$ is enforced by the delta function).

%%%%%%%%%%%%%%%%%%%%%%%%%%%%%%%%%%%%%%%%%%%%%% bare exchange
\subsubsection{Bare exchange  $\Sigma\_x$  }

We begin with the simplest bare exchange term.  We desire a matrix element of $\Sigma_x$ between two Bloch states $nk$ and $n'k$, and Bloch periodicity allows decomposition of this element into a sum over separate momentum transfers $q$:
\[
\bra{nk} \Sigma_x \ket{n'k}  = \frac{1}{N_k}\sum_q \bra{nk} \Sigma_x \ket{n'k}^q\,.
\]
For each $q$, we have
\[
\begin{aligned}
\bra{nk} \Sigma_x \ket{n'k}^q = -\sum_{v} \sum_g  \int dx \int dx'\ V_c(q+g) \ e^{-ig\cdot r}  {u_{nk}(x)}^*\\\times u_{vk+q}(x){u_{vk+q}(x')}^* {u_{n'k}(x')} e^{ig\cdot r'}\,.
\end{aligned}
\]
We now compare two different approaches to computing this matrix element.

First, we rearrange to make the outermost (final) sum over $g$ space:
\begin{multline*}
\bra{nk} \Sigma_x \ket{n'k}^q = -\sum_g V_c(q+G)\sum_{v}  \left[ \int dx\  e^{-ig\cdot r}  {u_{nk}(x)}^*u_{vk+q}(x) \right] \\\times \left[   \int dx'\  {u_{vk+q}(x')}^* u_{n'k}(x') e^{ig\cdot r'}\right] = -\sum_g V_c(q+g)\sum_{v}  \tilde{f}_g^{kqnv} \tilde{f}{_{g}^{kqn'v}}^*
\end{multline*}
where
\[
\tilde{f}_g^{kqnv} = IFFT[u_{nk}(x)^*u_{vk+q}(x)]\,.
\]
If $N_g$ represents the number of $g$ vectors, $N_r$ the size of the real-space grid, and $N_n$ the number of bands ($n$ and $n'$ indices combined) for which we wish to calculate $\tilde f_g^{kqnv}$, then computing and tabulating $\tilde f_g^{kqnv}$ requires $100N_r\ln N_r\cdot N_v\cdot N_n$ operations (where, again, an FFT costs $\approx 100N_r\ln N_r$ operations).  The computation of all the matrix elements requires $N_gN_n^2N_v$ operations.  Table~\ref{tab:ncompbareex} summarizes the main computational costs.

A second approach is to work in real-space, rewriting the matrix element as
\[
\begin{aligned}
\bra{nk} \Sigma_x \ket{n'k}^q = -   \int dx\ {u_{nk}(x)}^* \int dx'\ \left[\sum_g V_c(q+g) \ e^{-ig\cdot(r-r')}\right]   \\ \times \left[\sum_{v} u_{vk+q}(x) {u_{vk+q}(x')}^*\right] u_{n'k}(x')  \,.
\end{aligned}
\]
which, broken into stages, starts with computing the two separate matrices
\[
B^{k+q}_{x, x'} = \sum_{v} u_{vk+q}(x) u_{vk+q}(x')^*  \ \ , \ \  \mathcal{V}^q(r,r')=\sum_g V_c(q+g) e^{-ig\cdot (r-r')} 
\]
which are multiplied entry by entry to produce
\[
C^{kq}_{x,x'} = B^{k+q}(x,x')\mathcal{V}^q(r,r')\,.
\]
We then perform matrix vector operations to find the vectors
\[
J_{x,n'}^{kq} = \int dx'\ C^{kq}_{x,x'} u_{n'k}(x').
\]
The matrix elements are then computed by the dot products
\[
\bra{nk} \Sigma_x \ket{n'k}^q = - \int dx\, u_{nk}(x)^* J_{x,n'}^{kq}.
\]
Table~\ref{tab:ncompbareex} summarizes the computational costs of each part.

\begin{table}
\begin{center}
\begin{tabular}{| c | c | c |}
\hline
{Approach} & {Task} & {Operation Count} \\
\hline
\multirow{2}{*}{$g$-space} & Compute $\tilde{f}_g^{kqnv}$ & $ 100 N_r\ln N_r\cdot N_vN_n$ \\
& $\sum_g  V_c(q+g) \sum_v \tilde{f}_g^{kqnv}\ \tilde{f}{_{g}^{kqn'v}}^*$ & $N_gN_vN_n^2$  \\ 
\hline
\multirow{5}{*}{$r$-space} & Compute $B^{k+q}$ &  $N_r^2N_v$ \\
& Compute $\mathcal{V}^q$ & $ 100N_r\ln N_r + N_r^2$\\
& Compute $C^{kq}$ & $N_r^2$\\
& Compute $J^{kq}$ & $N_r^2N_n$\\
& Compute $u^*\cdot J$ & $N_rN_n^2$\\
\hline
\end{tabular}
\caption{Operation counts for computing the bare exchange term in the GW self-energy.  The operation counts are for a single $q$ momentum transfer and a single $k$ point.  A complex FFT is taken to cost $100N_r\ln N_r$ operations.}
\label{tab:ncompbareex}
\end{center}
\end{table}

% \begin{table}
% \begin{center}
% \begin{tabular}{ c | c  || c | c   }
% \multicolumn{2}{ c |}{$g$-space }  & \multicolumn{2}{| c }{$r$-space}  \\ 
% Task & Operation count & Task & Operation count\\
% \hline
% Compute $\tilde{f}_g^{kqnv}$  & $ 100 N_r\ln N_r\cdot N_vN_n$ 
% &Compute $B^{k+q}$ &  $ N_r^2N_v$  \\
% \hline
% $\sum_g  V_c(q+g) \sum_v \tilde{f}_g^{kqnv}\ \tilde{f}{_{g}^{kqn'v}}^*$ & $N_gN_vN_n^2$ 
% &Compute $\mathcal{V}_c$ &  $ 100N_r\ln N_r + N_r^2$  \\
% \hline
% & &  Compute $C^{k+q}$  & $ N_r^2$ \\ \hline
% & & Compute $J^{kq} $ & $N_r^2N_n$\\ \hline
% & & Compute $ u^* \cdot J$ & $N_rN_n^2$\\
% \end{tabular}
% \end{center}
% \end{table}

Comparing the $g$-space and $r$-space methods  requires some simple estimates of the relative sizes of the various parameters.  The largest parameters are $N_r\approx 2N_g$ which are the finest level of description in the physical problem.  The next largest number is $N_v$ which is the number of occupied states, and a well converged calculation has $N_r,N_g\gg N_v$: we require many plane waves or grid points per electronic state.  A typical ratio may be $N_r/N_v=500$.  Next, $N_n\ll N_v$ since typically we only inquire about the GW corrections for a limited subset of electronic states around the Fermi level. A typical ratio may be $N_n/N_v=0.2$.

With these numbers in mind, the $g$-space method has cubic and quartic scaling parts, the crossover happens approximately when $N_v\approx 1000$.  On the other hand, the $r$-space method is fundamentally cubic scaling and is dominated by the computation of $B^{k+q}$ which scales as $N_r^2N_v$.  Matching the cubic $r$-space operation count to the quartic one leads to a crossing at $N_v\approx$ 25,000.  This number of electronic states is so large that, even with growth of computer power in the near term, we do not expect to reach this size calculation on a routine basis without breakthrough reduced order methodology.  Hence, we opt for the $g$-space method for the computation of bare exchange.

%%%%%%%%%%%%%%%%%%%%%%%%%%%%%%%%%%%%%%%%%%%%%% Screened exchange
\subsubsection{Static screened exchange $\Sigma_{sex}$}
Next we discuss the static screen exchange part of the GW-self energy. This  $\Sigma_{sex}$ is described by: 
\begin{equation}
\begin{aligned}
\bra{nk} \Sigma_{sex} \ket{n'k}^q = -\sum_{v} \sum_{g,g'}\ S_{g,g'}^q  \int dx \int dx' \ e^{-ig\cdot r}  {u_{nk}(x)}^*\\\times u_{vk+q}(x){u_{vk+q}(x')}^* {u_{n'k}(x')} e^{ig\cdot r'}
\end{aligned}
\end{equation}
where
\[
S_{g,g'}^q  =  \sqrt{V_c(q+g)}[(\epsilon^q)^{-1}_{g,g'} -\delta_{g,g'}]\sqrt{V_c(q+g')}\,.
\]
Now we compare different approaches to computing the above matrix element.  
First, we start with a conventional $g$-approach where we use the same $\tilde f_g^{kqnv}$ above.  The matrix element then turns into
\[
\bra{nk} \Sigma_{sex} \ket{n'k}^q = -\sum_{gg'}  \sum_{v}\ \tilde{f}_g^{kqnv}\ S_{gg'}^q \ \tilde{f}{_{g'}^{kqn'v}}^*
=\sum_g \sum_v  \tilde{f}_g^{kqnv} \ T^{kqn'v}_g
\]
where 
\[
T^{kqn'v}_g  =  \sum_{g'} S_{gg'}^q \ \tilde{f}{_{g'}^{kqn'v}}^*\,.
\]
Computing $T$ is the most expensive 
term in this approach.
Table \ref{tab:ncompsex} summarizes the main computational costs. 

A second approach is to work in real space. Similar to the bare exchange, we first form the matrix $B^{k+q}_{x,x'}$ described above.  We then write the matrix elements in terms of $B^{kq}$ and integrals over r-space:
\[
\bra{nk} \Sigma_{sex} \ket{n'k}^q = - \int dx \int dx' u_{nk}(x)^*  u_{n'k}(x')  B^{k+q}_{x,x'} \sum_{g,g'}S_{g,g'}  e^{-ig\cdot r}   e^{ig'\cdot r'}
\]
We reorganize this by defining the matrix $\bar S^q$ as
\[
\bar{S}^q_{r,r'} = \sum_{g,g'}S^q_{g,g'}  e^{-ig\cdot r}   e^{ig'\cdot r'}
\]
which is computed via column-wise and then row-wise FFTs.  We then perform a point-wise multiplication (entry by entry) to create the matrix $R^{kq}_{x,x'}=B^{k+q}_{x,x'}\bar S^q_{r,r'}$. The matrix element then becomes
\[
\bra{nk} \Sigma_{sx} \ket{n'k}^q = - \int dx\  u_{nk}(x)^*  \int dx' u_{n'k}(x') R^{kq}_{x,x'} \\
\]
This is most effectively computed by first forming the matrix-vector product
\[
K^{kq}_{xn'} = \int dx'\ R^{kq}_{x,x'}u_{n'k}(x')
\]
and then overlap integrals for the final matrix element:
\[
\bra{nk} \Sigma_{sex} \ket{n'k}^q = - \int dx\  u_{nk}(x)^* K^{kq}_{xn'}
\]
See Table \ref{tab:ncompsex} for the computational cost for each part of this $r$-space method.

A third approach is a combination of $g$-space and  the eigen-representation of $S_{g,g'}^q$. 
In this approach, we use the diagonal form of $S^q$,
\[
S^q_{g,g'} = \sum_{\alpha}A_{g,\alpha}^q \cdot \sigma^q_{\alpha} \cdot {A_{g',\alpha}^q}^*\,.
\]
We insert this into our $g$-space expression from above:
\[
\begin{aligned}
\bra{nk} \Sigma_{sex} \ket{n'k}^q = -\sum_{gg'}  \sum_{v}\ \tilde{f}_g^{kqnv}\ S_{gg'}^q \ \tilde{f}{_{g'}^{kqn'v}}^* \\ = -\sum_{g,g',v,\alpha} A_{g,\alpha}^q \cdot \sigma^q_{\alpha} \cdot {A_{\alpha,g'}^q}^*  \cdot \tilde f^{kqnv}_g \cdot \tilde{f}{_{g'}^{kqn'v}}^*\,.
\end{aligned}
\]
For efficiency, we first create $\zeta$ as
\[
\zeta^{nv}_\alpha = \sum_g \tilde f_g^{kqnv} \, A_{g,\alpha}^q 
\]
and the use it to compute the matrix element
\[
\bra{nk} \Sigma_{sex} \ket{n'k}^q = -\sum_{\alpha} \sigma^q_{\alpha} \sum_{v} \zeta^{nv}_\alpha {\zeta^{n'v}_\alpha}^*
\]
If the number of eigenmode $\alpha$ needed for convergence $N_\alpha$ is  small, this approach provides significant computational savings.
The number of computation in this approach is  tabulated in  Table~\ref{tab:ncompsex}.

A final variant is to combine $r$-space and  eigen-representation of $S^q$.  Combining this eigen-representation with $B^{k+q}_{x,x'}$, we have 
\[
\begin{aligned}
\bra{nk} \Sigma_{sex} \ket{n'k}^q = - \int dx \int dx'\ u_{nk}(x)^* u_{n'k}(x') \\ \times \sum_{g,g',\alpha}A_{g,\alpha}^q \cdot \sigma^q_{\alpha} \cdot {A_{\alpha,g'}^q}^*\cdot  e^{-ig\cdot r}   e^{ig'\cdot r'} \cdot B^{k+q}_{x,x'}
\end{aligned}
\]
To compute this efficiently, first perform FFTs of $V_{\alpha, g}^q$ to get them in the $r$-space,
\[
U_{r,\alpha}^q = \sum_g e^{-ig\cdot r}A_{g,\alpha}^q\,.
\]
We sum over $\alpha$ while doing an entry-by-entry multiply to find $\tilde B^{kq}_{x,x'}$:
\[
\tilde{B}^{kq}_{x,x'} = \sum_{\alpha} U^q_{r,\alpha} \cdot \sigma^q_{\alpha} \cdot {U^q_{r',\alpha}}^* \cdot B^{k+q}_{x,x'} \,.
\]
This is followed by matrix-vector multiplies
\[
L^{kq}_{x,n'} = \int dx' \ \tilde{B}^{kq}_{x,x'}u_{n'k}(x')
\]
and dot products
\[
\bra{nk} \Sigma_{sex} \ket{n'k}^q= - \int dx\ 
u_{nk}(x)^*  L_{x,n'}^{kq}.
\]

\begin{table}
\begin{center}
\begin{tabular}{| c | c | c | r | r | r| r| }
\hline
%\multicolumn{2}{| c }{$g$-spcae }  & \multicolumn{2}{| c }{$r$-space} & \multicolumn{2}{| c |}{r-space matrix} \\ 
{Approach }  & {Task } & {Operation Count} \\ 
% & \multicolumn{2}{| c |}{Approach III}\\ 
%\multicolumn{1}{ c }{Steps } & \multicolumn{1}{| c }{Approach I}  \\ 
\hline
 \multirow{3}{*}{$g$-space} & Compute $\tilde{f}_g^{kqnv}$ & $  N_vN_n\cdot 100 N_r \ln N_r$ \\
%\hline
&  Compute $T_g^{kqn'v} $ & $N_vN_nN_g^2$ \\
&  $\sum \tilde{f}_g^{kqnv} T_g^{kqn'v} $ & $N_vN_n^2N_g $ \\
%&  $\sum_g \ v^q(g)\ \sum_l \tilde{f}_g^{nl}$ $\tilde{f}{_{g}^{n'l}}^*$ & $ 1.6\times N_L^4$ \\
\hline
\multirow{5}{*}{$r$-space} & Compute $B^{k+q}_{x,x'}$ & $N_r^2N_v$ \\
& Compute $\bar S^q_{r,r'}$ & $2N_r^2\cdot 100\ln N_r$  \\
& Compute $R^{k,q}$ &$N_r^2 $ \\
& Compute $K^{k,q}$ & $N_nN_r^2$ \\
& Compute $ u^{*}\cdot K$ & $N_n^2N_r$ \\
\hline
\multirow{3}{*}{$g$-space+eigen} & Compute $\tilde{f}_g^{kqnv}$ & $ N_vN_n \cdot 100N_r \ln N_r$ \\
& Compute $\zeta^{nv}_\alpha$ &  $N_vN_nN_gN_\alpha$ \\
&  $\sum \sigma_\alpha \zeta^{nv}_\alpha{\zeta^{n'v}_\alpha}^*$ & $N_vN_n^2N_\alpha $\\
\hline
\multirow{3}{*}{$r$-space+eigen} & Compute  $B^{k+q}$ & $N_vN_r^2$ \\
& Compute $U^q$ & $N_\alpha \cdot 100N_r\ln N_r$  \\
& Compute $\tilde B^{kq}$ & $N_\alpha N_r^2$\\
\hline
\end{tabular}
\caption{Operation counts for computing the static screened exchange term of the static COHSEX GW-self energy based on a variety of methods (see text). For simplicity, the table shows operation counts for a single $k$-point.}
\label{tab:ncompsex}
\end{center}
\end{table}

 We now compare the four methods using numerical estimates for the relations of the various parameters from the above section on bare exchange. We again use a typical ratio of $N_r/N_v=500$. We will use $N_v>100$ as a reasonable definition of a ``large''
system; $N_\alpha/N_g$ is assumed small but larger than $10^{-2}$.  
The worst scaling parts for each calculation are the calculations of $T$, $B^{k+q}$,  $\zeta$,
and  ${\tilde{B}}^{kq}$ for the $g$-space, $r$-space, $g$-space+eigen and the 
$r$-space+eigen methods, respectively. The scalings for the respective terms are  $3\times 10^3\cdot N_v^4$, 
$10^5 \cdot N_v^3$, $ (N_\alpha/N_g)\cdot 6\times10^3 \cdot N_v^4 $, and 
$(N_{\alpha}/N_g) \cdot 2 \times 10^7\cdot N_v^3$.   Comparing the $r$-space and $g$-space methods, the $r$-space has a smaller number of computations when $N_v>100$.  The $r$-space methods without and with eigen-representation are comparable, with the eigen approach being more expensive computationally once $N_\alpha/N_g>10^{-2}$.  Comparing the quartic scaling $g$-space+eigen method with the cubic scaling $r$-space method, the cross over occurs when $N_v\approx (1/6)/(N_\alpha/N_g)<1700$.

This scaling analysis indicates that as far as the screened exchange is concerned, the $r$-space method is the winner for large systems with the $g$-space method coming second. However, we will not opt for the $r$-space approach since the most expensive part of the GW-self energy, the dynamical screened Coulomb hole part (see the next section), becomes extremely costly in the $r$-space approach.  Instead we choose the $g$-space approach which is somewhat suboptimal for static screened exchange but saves a great deal of computation for the dynamic self-energy.  For completeness and as a correctness test, we compare the convergence of the static self-energy between our $g$-space approach and that of the BerkeleyGW software~\cite{BGW} in Figure~\ref{fig:cohsexconvergence}: since both are $g$-space methods, their convergence rates are quite similar.

\subsubsection{Static Coulomb hole $\Sigma_{coh}$}

\begin{figure}
\centering
\subcaptionbox{static screened exchange}{
\includegraphics[width=0.8\textwidth]{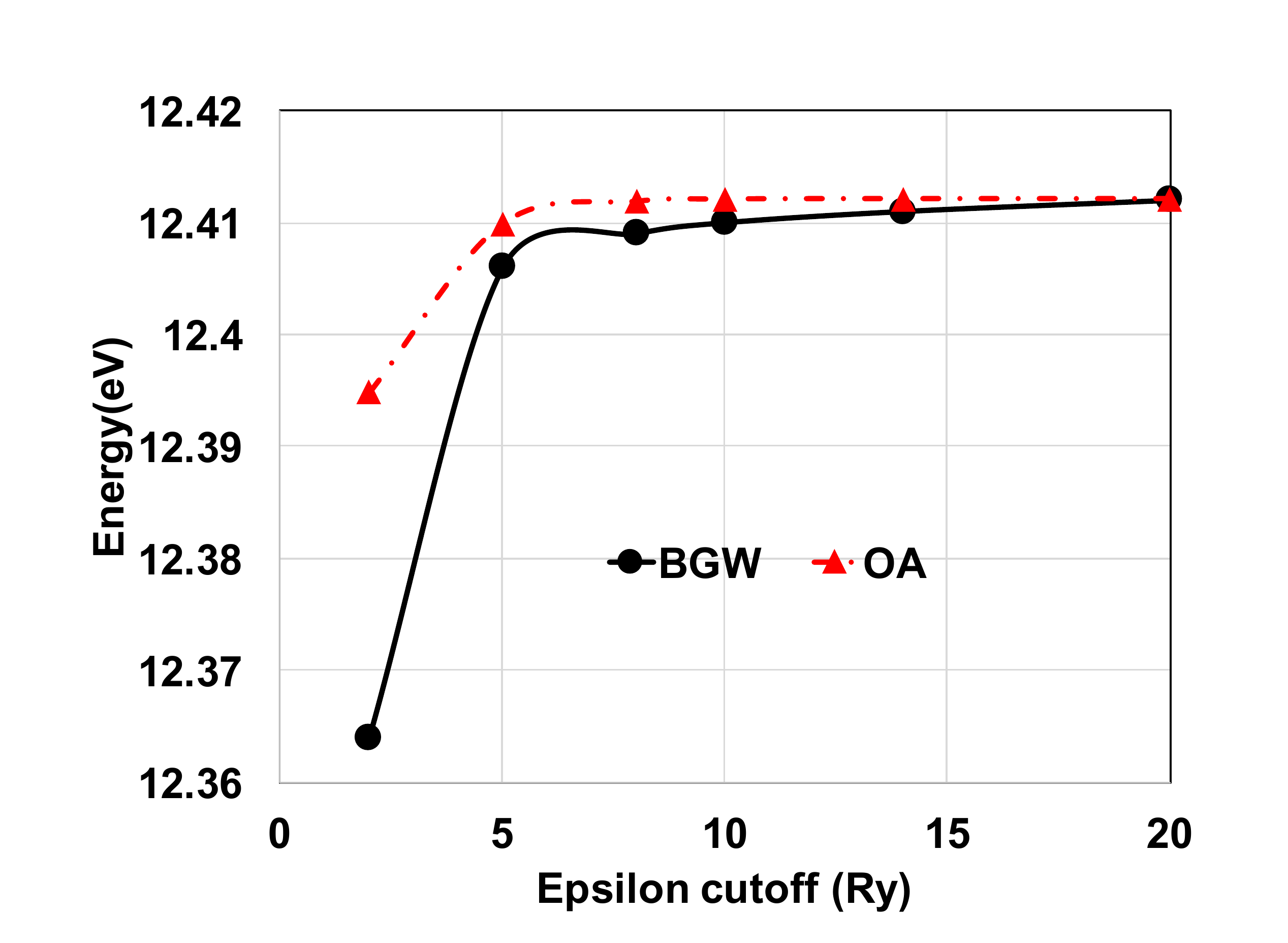}
}
\subcaptionbox{static Coloumb-hole part of the COHSEX self-energy}{
\includegraphics[width=0.8\textwidth]{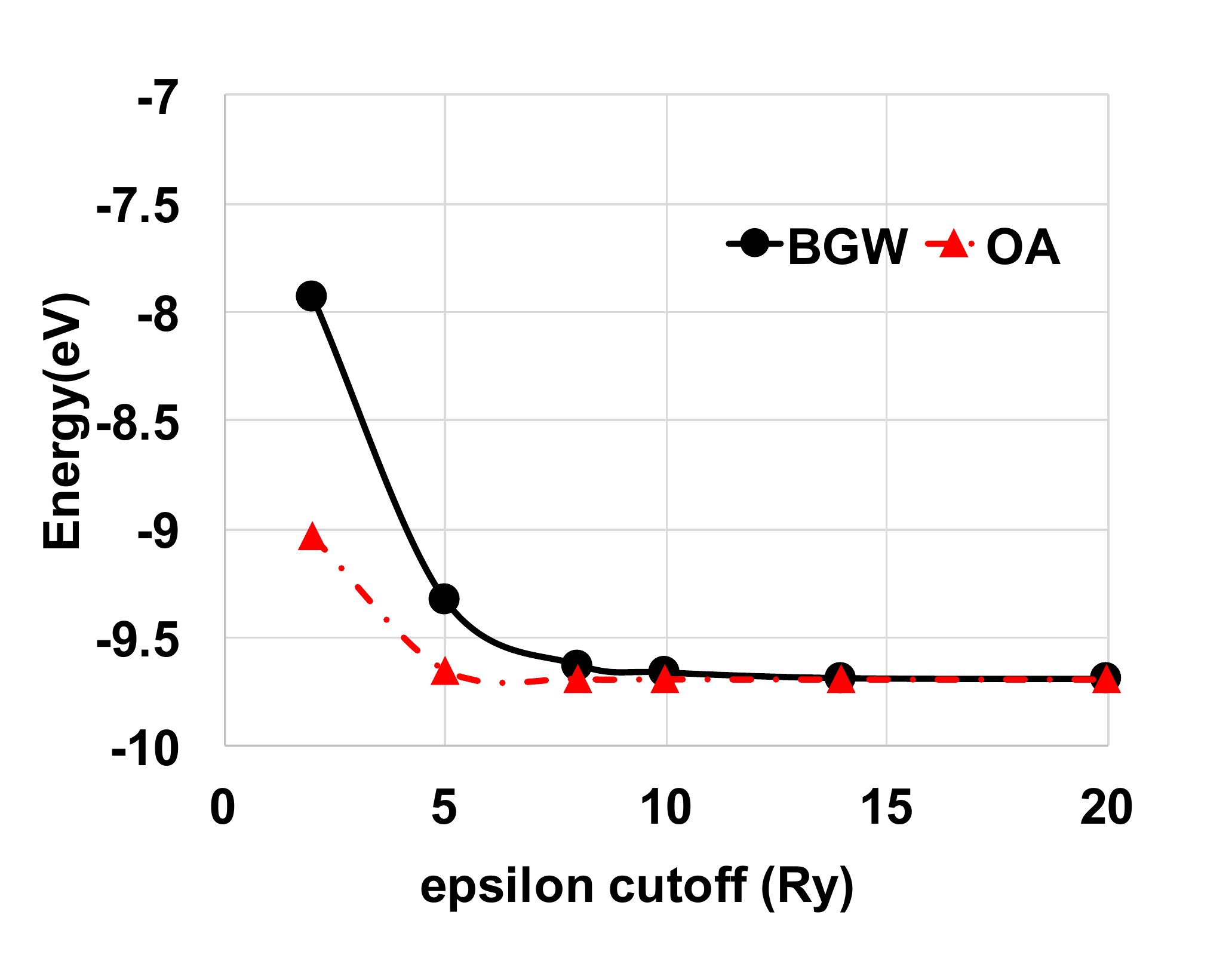}
}
\caption{Convergence tests as a function of the polarizability matrix cutoff for bulk Si.
 Comparison with BerkeleyGW (BGW)~\cite{BGW} software (black line) yields a very similar trend since both
OpenAtom (OA) and BGW use a reciprocal ($g$-space) approach.}
\label{fig:cohsexconvergence}
\end{figure}

The last term in the GW-self energy is the Coulomb-hole part.  Within COHSEX, this term is not computationally expensive compared to other two terms, so we simply opt for a standard $g$-space approach:
\begin{eqnarray*}
\bra{nk} \Sigma_{coh}\ket{n'k} & = & -\frac{1}{2N_k}\sum_{g,g',q}  \int dx\  {u_{nk}(x)}^*\ u_{n'k}(x)\,S^{q}_{g,g'}e^{i(g-g').r}\\
 & = & -\frac{1}{2}\sum_{g,g',q} S^{q}_{g,g'} Y^{knn'}(g'-g)
\end{eqnarray*}
where $Y$ is the FFT of the product of the two periodic wave functions (very similar to $\tilde f$ above):
\[
Y^{knn'}(g) = \int dx\  {u_{nk}(x)}^*\ u_{n'k}(x)e^{-ig\cdot r}.
\]
The convergence of this term is shown in Figure~\ref{fig:cohsexconvergence}.

%\end{document}

\subsection{Dynamic self-energy}\label{sec:dynamicsigma}
%%\input{dynsigma}

%%%%%%%%%%%%%%%%%%%%%%%%%%%%%%%%%%%%%%%%%%%%%%
% Dynamic sigma
%%%%%%%%%%%%%%%%%%%%%%%%%%%%%%%%%%%%%%%%%%%%%%

We use the plasmon-pole (PP) form of Eq.~(\ref{eq:eigdecomp}), (\ref{eq:ppform}), and (\ref{eq:ppfreq}) from Section \ref{sec:pp} to write the following expressions for the dynamical screened exchange matrix element   
\begin{multline}
\bra{nk} \Sigma_{sex}(\omega) \ket{n'k}^q =
- \sum_{v,g,g',\alpha} \!\int \!\! dx \!\! \int \!\! dx'\,  e^{-ig\cdot r} A^q_{g,\alpha} {A^{q}_{g',\alpha}}^* e^{ig'\cdot r'} \\ \times  u_{nk}(x)^*u_{vk+q}(x) u_{vk+q}(x')^* u_{n'k}(x')     \, 
\frac{\sigma^q_\alpha {\omega_\alpha^q}^2 }{(\omega^q_\alpha)^2 - (\omega-E_v^{k+q})^2}
\label{eq:dynsx}
\end{multline}
while the dynamic Coulomb-hole matrix element is
\begin{multline}
\bra{nk} \Sigma_{coh}(\omega) \ket{n'k}^q = \sum_{b,g,g',\alpha}\!\int \!\! dx \!\!  \int \!\! dx'\, e^{-ig\cdot r} A^q_{g,\alpha} {A^{q}_{g',\alpha}}^* e^{ig'\cdot r'} \\\times  u_{nk}(x)^*u_{bk+q}(x) u_{bk+q}(x')^* u_{n'k}(x')    \,
\frac{\sigma^q_\alpha \omega_\alpha^q}{2(\omega^q_\alpha + E_b^{k+q} - \omega)}
\label{eq:dyncoh}
\end{multline}
where the index $b$ sums over all available  $N_b=N_v+N_c$ bands (occupied and unoccupied).
Above two equations have very similar structure except the slightly different energy denominator.
As $N_c\gg N_v$ for converged GW calculations, the Coulomb-hole calculation will always be more expensive. Thus, we will only analyze the Coulomb-hole case to understand scalings and draw our conclusions on the best implementation.

The first approach is $r$-space.  As above, we FFT the $A^q_{g,\alpha}$ into the $U^q_{r,\alpha}$.  We then form the matrix 
\[
D^{kq}_{x,x'} = \sum_{\alpha} U_{r,\alpha}^q {U_{r',\alpha}^{q}}^*\sum_{b} \frac{ \sigma^q_\alpha \omega_\alpha^q}{2(\omega^q_\alpha + E_b^{k+q} - \omega)} u_{bk+q}(x)u_{bk+q}(x')^*
\]
and compute the matrix element
\[
\bra{nk} \Sigma_{ch}(\omega) \ket{n'k}^q = - \int dx \ u_{nk}(x)^* X^{kq}(x,n')
\]
where
\[
X^{kq}_{x,n'} = \int dx'\ D^{kq}_{x,x'} u_{n'k}(x')
\]

The second approach is calculating the matrix element in $g$-space. We first combine multiple energy-dependent quantities into $\Delta$:
\[ 
\Delta^{b}_\alpha= \frac{\sigma^q_\alpha \omega_\alpha^q}{2(\omega^q_\alpha + E_b^{k+q} - \omega)}.
\]
We then combine $V$ with $\tilde f$ to form
\[
\zeta^{nb}_\alpha = \sum_g A_{g,\alpha}^q \tilde f^{kqnb}_g\,.
\]
The final form of Coulomb-hole is
\[
\bra{nk} \Sigma_{ch}(\omega) \ket{n'k}^q =\sum_\alpha \sum_{b} \Delta^b_\alpha \zeta^{nb}_\alpha (\zeta^{n'b}_\alpha)^*\,.
\]
%%% table %%%
\begin{table}
\begin{center}
\begin{tabular}{| c | c | c |}
\hline
{Approach} & {Task} & {Operation Count}\\
\hline

\multirow{4}{*}{$r$-space } & Compute $U^q$ & $N_{\alpha} \cdot 100 N_r\ln N_r$  \\
& Compute $D^{kq}$ & $N_{\alpha} N_b N_r^2$ \\
&Compute $X^{kq}$ & $N_r^2N_n$ \\
& Compute $u\cdot X$ & $N_rN_n^2$ \\

\hline

\multirow{4}{*}{$g$-space } & Compute $\Delta^b_\alpha$ & $N_{\alpha}  N_b$ \\ 
& Compute $\tilde f^{kqnb}_g$ & $N_nN_b\cdot 100N_r\ln N_r$\\
& Compute $\zeta^{nb}_\alpha$ & $N_{\alpha}N_gN_nN_b$ \\
& Compute $\sum \Delta^b_\alpha \zeta^{nb}_\alpha (\zeta^{n'b}_\alpha)^*$ & $N_{\alpha} N_n^2 N_b$\\

\hline
% first line
%Compute $U^q$ & $N_{\alpha} \cdot 100 N_r\ln N_r$ & $\Delta^b_\alpha$ & $N_{\alpha}  N_b$ \\ 
% second line
%Compute $D^{kq}$ & $N_{\alpha} N_b N_r^2$ & $\tilde f^{kqnb}_g$ & $N_nN_b\cdot 100N_r\ln N_r$\\
% third line
%Compute $X^{kq}$ & $N_r^2N_n$ & $\zeta^{nb}_\alpha$ & $N_{\alpha}N_gN_nN_b$ \\ 
% 4-th line
%Compute $u\cdot X$ & $N_rN_n^2$ & $\sum \Delta^b_\alpha \zeta^{nb}_\alpha (\zeta^{n'b}_\alpha)^*$ & $N_{\alpha} N_n^2 N_b$\\
\end{tabular}
\caption{Operation counts for computing the dynamic Coulomb-hole GW self-energy for real and reciprocal space based methods. For simplicity, the table shows operation counts for a single $k$-point.}
\label{tab:dynsig}
\end{center}
\end{table}

To compare the scaling and operation counts for the $r$-space and $g$-space methods, we indicate the intermediate steps and scalings in Table~\ref{tab:dynsig}.
As $N_\alpha$, $N_b$, $N_n$, $N_r$, and $N_g$ are all proportional to the size of the system (number of atoms $N$), both approaches scale as $N^4$. However, the pre-factor of $r$-space method is especially large due to the calculation of the $D^{kq}$ matrix.  Hence, unlike the computation of $P$, the $r$-space method is not competitive for the computation of the self-energy.
For any system size, $g$-space method  always wins the scaling battle. Our implementation of dynamic sigma is thus based on the $g$-space method. 

\begin{figure}
\centering
\includegraphics[width=0.65\textwidth]{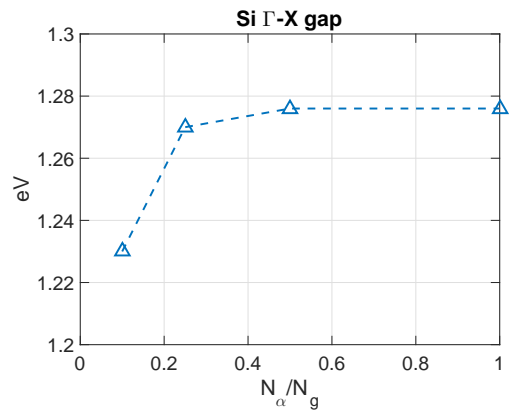}
\caption{$\Gamma$-$X$ energy gap for Si using the dynamic GW self-energy method. The horizontal axis is the fraction of plasmon-pole modes included in the summations for the self-energy.  A 2-atom cell with 8 k points was used. The energy cutoff for the polarizability matrix, and hence $\epsilon$,  was set to 10 Ry which translates into $N_g$ being 137 for $\Gamma$ and 150 for $X$.}
\label{fig:gap}
\end{figure}

Since we have observed the rapid convergence of $\epsilon(\omega)^{-1}$ within our PP model, we expect the dynamic GW self-energy to converge quickly as well with respect to the number of plasmon modes ($N_\alpha$) included in the calculation. This can greatly reduce the total number of computations needed for reasonable convergence levels.  For example, we varied the number of plasmon modes $N_\alpha$ included in the dynamic self-energy from 0.1$N_g$ to $N_g$, and calculated the occupied-unoccupied $\Gamma$-$X$ energy gap of silicon. Figure~\ref{fig:gap} shows the convergence of this enegy gap with respect to $N_\alpha$. The energy cutoff for the screening quantities is 10~Ry. Even with only 10\% of the plasmon modes, the gap is less than 0.05 eV from the final answer. Using 25\% of plasmon modes, $\Gamma$-$X$ gap is within 6 meV of the final answer. Hence, truncation of the high energy vdHL PP modes in the dynamic sigma can provide significant computational savings.

\section{Large scale parallelization of GW under charm++}\label{sec:scaling}

We now describe our approach to parallelization and then show performance results for the resulting software.  Our main focus is on the computation of the polarizability $P$ as that is the primary bottleneck of any GW calculation.

%\subsection{Parallelization using Charm++}

\subsection{Charm++ Parallel Runtime System}

\begin{figure}
\centering
\includegraphics[width=0.9\textwidth]{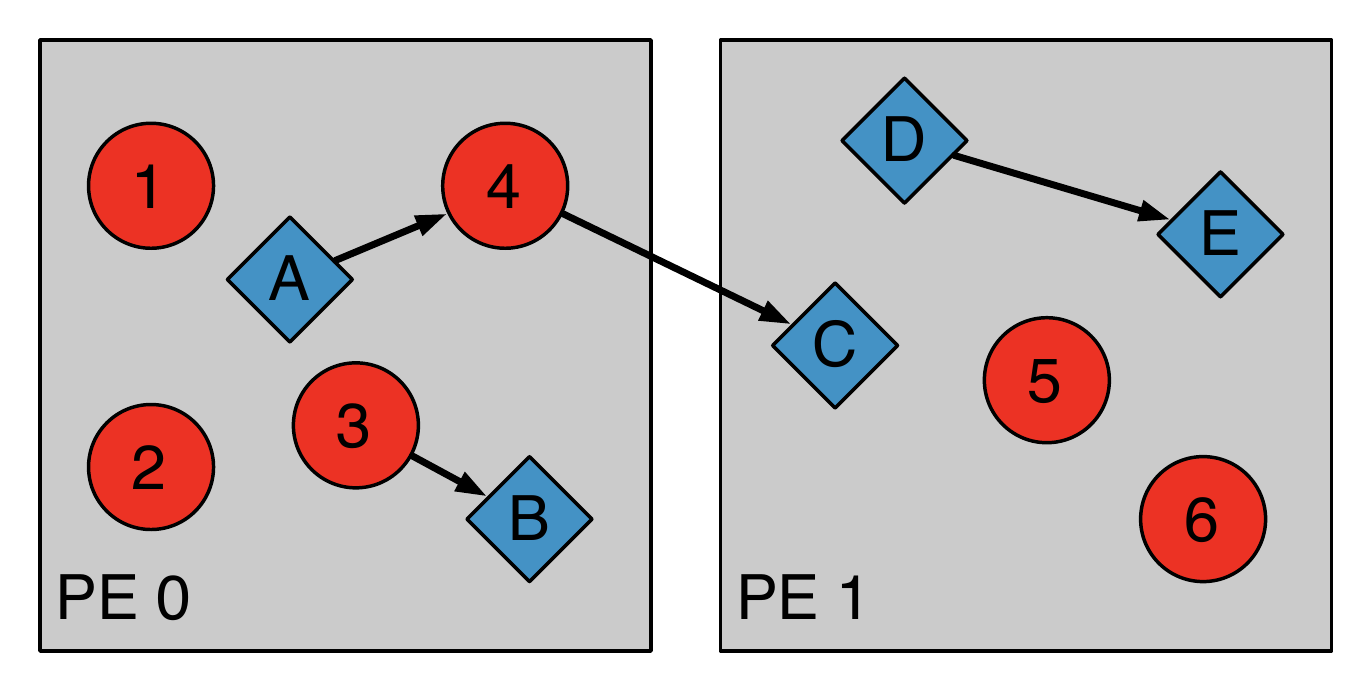}
\caption{A representation of an arbitrary Charm++ application comprised of two different collections of chares: one is shown as red circles and the other chare collection as blue diamonds.  The chares are mapped by the runtime system to two different processing elements (PE) (typically a physical core or a node). Arrows represent messages sent between chares.}
\label{fig:charm_model}
\end{figure}

The \openatom{} version of the GW computation is built on top of \charmpp{} \cite{charm_2014}, a parallel programming framework that utilizes object-based process virtualization. Charm applications are comprised of collections of parallel objects, called ``chares", managed by an adaptive runtime system. An application can define as many types of chares as is needed, and instantiate chares in collections of whatever size and dimensionality match the particular problem they are trying to solve. A schematic illustration of this {\em object-based decomposition} can be seen in figure~\ref{fig:charm_model}, where there are two different collections of chares represented here by the diamonds and circles. Chares communicate with one another by sending messages, represented by the arrows between chares in the figure. The underlying \charmpp{} runtime system manages the actual locations of the chares on the physical hardware, as well as handling communication between chares. This includes handling both within node and between node communication. 

Adaptive Runtime system management of the chares provides efficient scheduling of computation and communication to optimize utilization of both the CPU and network resources. Communication between chares is asynchronous and one-sided which gives the runtime flexibility to schedule execution of chares as messages for them become available, achieving an adaptive overlap between communication and computation without requiring explicit direction from the application developer. The separation of application logic from hardware resources also allows independent and modular mapping of chares to hardware without affecting application correctness. This can have significant performance implications~\cite{jain:isc2016} where performance can be improved by up to 30\% using topology aware mapping schemes tailored to two different execution environments.

\subsection{Parallelization of GW}
We describe the parallelization of different phases of GW using \charmpp{} programming model. First we describe the parallelization of the polarizability matrix $P$ computation, as it is the primary computation phase affecting scaling and execution time. Next we describe the parallelization of subsequent phases to compute the dielectric $\epsilon$ matrices, $\epsilon^{-1}$ matrices and self-energy $\Sigma$ calculations.

\subsubsection{Parallel Caching and Computation for $P$}\label{problemDecomposition}
\begin{figure}
% * <lvk2408@gmail.com> 2018-07-13T04:00:31.430Z:
% 
% The labeling (that 0 1 2 and 3 are "steps" is not stated (neither in legend nor in text --- Sanjay)
% 
% ^.
\centering
\includegraphics[width=0.9\textwidth]{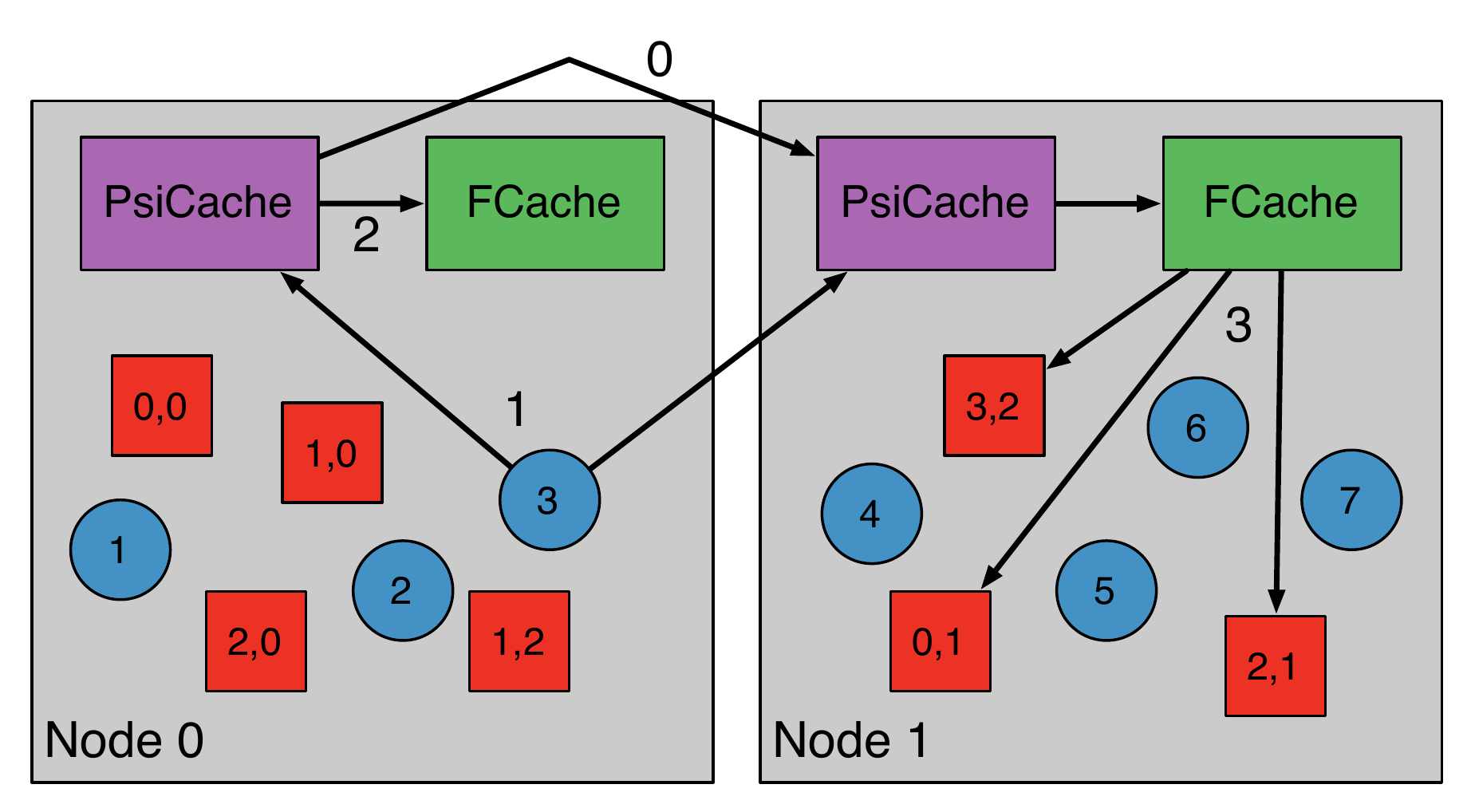}
\caption{The decomposition of the OpenAtom GW computation with Charm++ shown on two compute nodes. The PsiCache and FCache have one chare per node and store data that can be directly accessed from anywhere on the node. The state chares (blue) and matrix chares (red) are overdecomposed with many chares per node.}
\label{fig:gw_decomp}
\end{figure}
The parallel decomposition of the GW computation in \charmpp{} has two basic components: a caching structure for storing temporary data on each hardware computational node, and computational structures that store more persistent data and orchestrate the computation on that data. In this section we focus specifically on the computation of the polarizability matrix, $P$, due to the fact that it generally is the most computationally intensive part of the computation. The entire decomposition for this phase of the algorithm is shown in Figure~\ref{fig:gw_decomp}, and described in more detail below. Other phases are implemented by applying the same principles --- using multiple interacting collections of chares to send data and orchestrate the computation.

First, the caching structure is necessary due to the impossibility of storing data for all of the intermediate computations of large data sets on each node. The two different caches, \texttt{PsiCache} and \texttt{FCache}, are special chare collections that fix exactly one instance of each to a single hardware node. This allows all other chares on the node to access the cached data directly via memory pointers rather than via messaging. In the setup of the computation, each occupied state is duplicated across all \texttt{PsiCaches}, as represented by step 0 in figure~\ref{fig:gw_decomp}. During the formation of $P$, unoccupied states are broadcast to all nodes in a pipelined fashion (step 1 in figure~\ref{fig:gw_decomp}). Upon receiving an unoccupied state, the \texttt{PsiCache} multiplies it with all occupied states using \texttt{CkLoop}, a within-node parallelization construct in \charmpp{} similar to \texttt{OpenMP} \texttt{for} loops. This forms a set of $f$ vectors of Eq. (\ref{eq:fdef}) that are used to form the $P^q$ matrix (step 2 in figure~\ref{fig:gw_decomp}).

For the formation of the $P^q$ matrix, there are two chare collections used in the actual orchestration of the computation: state chares and matrix chares. The state chares are a one-dimensional array of chares that load and store the electronic states, and are represented in figure~\ref{fig:gw_decomp} as the blue circles. As stated in the previous paragraph, subsets of these chares broadcast their states to the \texttt{PsiCaches} on every node during step 1 of the computation so that each node can compute the next set of $f$ vectors. The other primary chare collection used during the $P^q$ matrix formation is the \texttt{PMatrix} chare. These chares are instantiated as a two-dimensional array of chares, as represented by the red squares in figure~\ref{fig:gw_decomp}, with each chare holding a tile of the $P^q$ matrix. The size of these tiles and, by association, the number of chares in the array is configurable at runtime. Once a set of $f$ vectors is computed, each matrix tile can directly access the portions of the $f$ vectors that correspond to its entries in the matrix, and perform the outerproduct with a BLAS matrix multiplication call (step 3 in figure~\ref{fig:gw_decomp}, only shown on node 1 for clarity). These computations can happen independently on each core of a node, and the work for these computations is automatically overlapped with work and communication done by the \texttt{PsiCache} and state chares. This allows for multiple unoccupied states to be in flight at once, and the different parts of the computation can overlap to fully utilize the CPU. Once all the matrix chares on a node have read the $f$ vectors from the cache, the unneeded vectors can be discarded and the next set of state chares broadcasts a new set of unoccupied state data to the \texttt{PsiCaches}. This repeats until every unoccupied state has been broadcast and the polarizability matrix is complete. The performance of this first phase of the computation is explored in the next section and compared to another existing GW implementation to demonstrate its scalability.

An additional benefit which can be exploited is the fact that the entries of the $f$ vectors that need to be computed on a given node depend on the tiles of the matrix that are mapped to that node. We improve scaling by reducing the computation of $f$ vectors on each node to only those entries of $f$ vectors required by the tiles of matrix mapped to the node. By developing a different mapping of \texttt{PMatrix} chares to nodes, it is possible to further minimize the computation done on each node, as well as the total amount of memory required for the temporary storage of $f$ vectors. This change  allows the computation to scale more effectively to higher node counts by reducing of the cost of the $f$ vector computation as the number of tiles per node is decreased in a strong scaling problem. This mapping can be developed independently of the rest of the application, as the mapping of chares is completely separate from their computation. Furthermore, this idea can also be applied to matrix computations in later phases of the algorithm as well.

\subsubsection{Parallel Caching and Computation for $(\epsilon^q)^{-1}$ and $\Sigma$} %todo Also need to include phase-2 somewhere
We describe the parallelization of $\epsilon^q$ matrix inversion and $\Sigma$ calculations that follow the computation of the static polarizability matrix $\tilde P^q$. These could affect scaling if $P$ is sufficiently well parallelized that we approach the Amdahl limit (however, we are far from this limit in all our examples and test cases).

For the formation of $\epsilon^q$ and $(\epsilon^q)^{-1}$, we have the \texttt{EpsilonMatrix} and\\ \texttt{EpsilonInverseMatrix} collection structures which are matrix chare arrays - similar to $P$. Similar to \texttt{PMatrix}, these are instantiated as two-dimensional arrays of chares. During $\epsilon^q$ matrix formation, each \texttt{PMatrix} chare applies the energy cutoff, that is cached in \texttt{PsiCaches} on every node, on its tile of the $\tilde P^q$ matrix. These smaller tiles on applying cutoff are sent to corresponding \texttt{EpsilonMatrix} chares. The resulting \texttt{EpsilonMatrix} dimensions are much smaller than that of the Polarizability matrix. Once each \texttt{EpsilonMatrix} chare has received its tile of the $\epsilon^q$ matrix, iterative matrix multiplication is performed on the \texttt{EpsilonMatrix} chares to compute \texttt{EpsilonInverseMatrix}. The distributed iterative matrix multiplication uses several instantiations of two-dimensional matrix chare arrays of the same dimension as \texttt{EpsilonMatrix} to represent intermediate matrices. Distributed matrix multiplication is performed by sending data from two collections of input matrix chares to output matrix chares that perform the matrix multiplication per tile. The resulting \texttt{EpsilonInverseMatrix} chares each hold a tile of the $(\epsilon^q)^{-1}$ matrix.

Static self-energy calculations are performed once \texttt{EpsilonInverseMatrix} has been computed. To compute the static self-energy, we need to assemble the bare exchange, screened exchange, and Coulomb hole contributions. Bare exchange calculations require only those $f$ vectors obtained by multiplying states from a user-provided list of bands (of length $N_n$). Similar to \texttt{PMatrix} formation, states from the band list are broadcast to all nodes and $f$ vectors are formed by multiplying these states with the all the occupied states duplicated during polarizability matrix computation across \texttt{PsiCaches} on all nodes. As before, $f$ vector computation is parallelized on each node using \texttt{CkLoop} constructs. Each set of $f$ vectors is stored in the \texttt{FCache} memory. In order to reduce the memory usage for caching the $f$ vectors, only portions of the $f$ vectors corresponding to the tiles of $(\epsilon^q)^{-1}$ matrix on each node are cached. Each \texttt{EpsilonInverseMatrix} chare then computes bare exchange values using $f$ vectors corresponding to its tile indices and contributes these values to a reduction across the \texttt{EpsilonInverseMatrix} chare array to obtain the bare exchange value for each $k$ point.

{\raggedright\setlength\parindent{0.5cm}
Similar to bare exchange computation, each \texttt{EpsilonInverseMatrix} chare computes screened exchange values. Each \texttt{EpsilonInverseMatrix} chare multiplies its tile of $(\epsilon^q)^{-1}$ matrix with corresponding tile of outer product of $f$ vectors. The resulting values are added and each element of the chare array collection \texttt{EpsilonInverseMatrix}  contributes to a reduction to produce the screened exchange value per $k$ point.
}

{\raggedright\setlength\parindent{0.5cm} 
Finally each \texttt{EpsilonInverseMatrix} chare computes Coulomb hole calculations by using $f$ vectors that are formed by dot product of only those states corresponding to the band list indices specified by the user. These states specified in the band list are cached in \texttt{PsiCaches} on each node during the broadcast of occupied and unoccupied electronic states by the state chares for the polarizability matrix calculations. Since there is a relatively small number of $f$ vectors, each \texttt{EpsilonInverseMatrix} chare first computes $f$ vectors from the cached states. Each \texttt{EpsilonInverseMatrix} chare then multiplies its own tile with corresponding $f$ vector values. Similar to bare and screened exchange, each chare then contributes its local sum to a reduction over the \texttt{EpsilonInverseMatrix} chare array collection to produce Coulomb hole values for $\Sigma$.
}

\subsection{Parallel performance}

We choose ``small",  ``medium'', and ``large''  sized systems for parallel testing.  They consist of  unit cells 
with 54, 108, and 432 atoms of Si describing the bulk diamond structure material.  The DFT description here is provided by the local density approximation (LDA) of Perdew-Zunger (PZ)~\cite{PZ} for exchange correlation together with a plane wave basis and norm conserving pseudopotentials~\cite{TM}.  The plane wave cutoff energy is 12 Ryd.  We used the Quantum Espresso software \cite{QE} to perform the DFT calculations and compute the band structure:  108 valence bands and 491 conduction bands are generated for 54 Si atoms unit cell. For ``medium'' sized systems, 216 valence and 2048 conduction bands are generated for Si; for ``large''  sized systems, we generated 648 valence and 10,000 conduction bands.  We then convert the wave function data to \openatom{} format by simple custom-built converters. We compute $P^q$ for one wave vector at $q=(0.0,0.0,0.001)$ (reciprocal lattice units).  

\begin{center}
\begin{figure*}
\centering
\subcaptionbox{Strong scaling results generating the same convergence level.  88$\times$88 chares are used by OA for these results.}{
\includegraphics[width=300pt, angle=0]{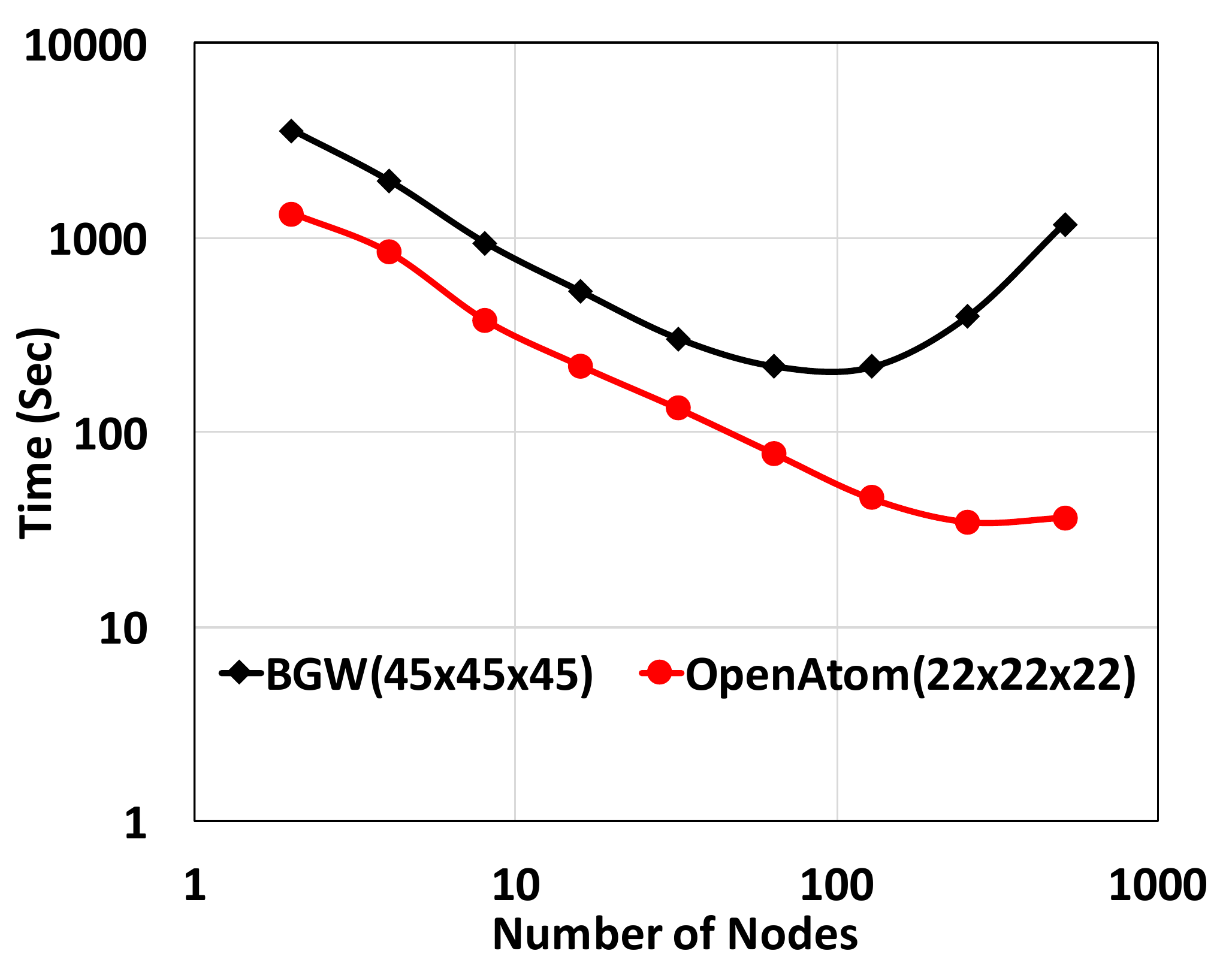}}
\subcaptionbox{Dependence of OA performance on the number of chares.}{
\includegraphics[width=300pt, angle=0]{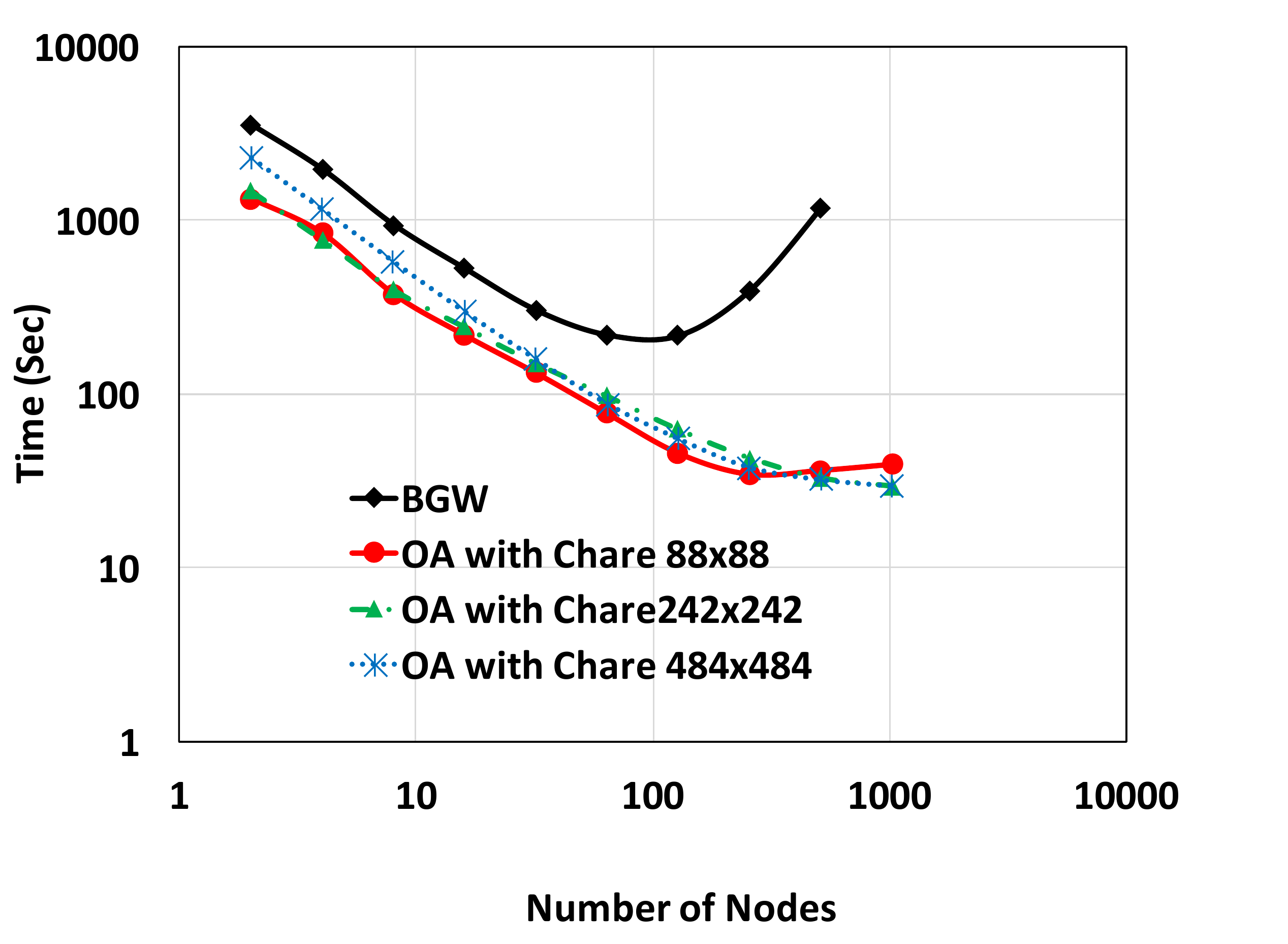}
}
\caption{ Performance scaling of the polarizability matrix calculation with the Berkeley GW (BGW) and OpenAtom (OA) software packages on IBM BlueGene.  The system is bulk Si with a 54 atom unit cell, 12 Ryd plane wave cutoff, and 10 Ryd cutoff for the polarizability.
}
\label{fig:parallelscaling}
\end{figure*} 
\end{center}

For the computation of $\tilde P^q(g,g')$, the plane wave cutoff defining the set of $g$ vectors is taken to be 10 Ryd (on the single particle electronic states).  We have varied the FFT grid until convergence is reached to below 0.01\% error for the matrix elements $\tilde P^q(g,g')$ compared to those of full FFT grid.  This translates into a modest $22\times22\times 22$ FFT grid for the system with 54 Si atoms (``small'' sized system).  Figure~\ref{fig:parallelscaling} presents strong parallel scaling performance for this physical system using the 10-petaflop IBM Blue Gene/Q supercomputer Mira at Argonne National Laboratories.  32 threads per node are used in this calculation.  We see that \openatom{} shows excellent scaling up to $\sim$500 nodes which is impressive for the modest size of this example. For ``medium'' and ``large'' sized system, we use both Mira as well as the 13-petaflop National Center for Supercomputing Applications (NCSA) Cray supercomputer Blue Waters at the University of Illinois at Urbana-Champaign. 

To gauge the performance results in more pragmatic terms, we have compared the parallel performance of our \openatom{} implementation to that of the open source BerkeleyGW (BGW) \cite{BGW} software package of version 1.2.  BGW is also a plane wave based  software for GW calculations using the sum-over-states method to compute $\tilde P^q$.  The main methodological difference is that BGW uses the $g$-space approach of Eq.~(\ref{eq:gpsaceP}) to compute $\tilde P^q_{g,g'}$ directly whereas \openatom{} uses an $r$-space approach to compute $P^q_{r,r'}$ and then converts that to $\tilde P^q_{g,g'}$ via FFTs.  Direct apples-to-apples comparison between the two software packages is thus not completely straightforward.  To simulate actual usage in scientific applications, we decided to require both applications to generate the same physical level of convergence.  This translates into a 45$\times$45$\times$45 FFT gird for BGW.   As shown in Figure~\ref{fig:parallelscaling}, BGW shows excellent scaling up to $\sim$100 nodes for this 54 atom problem and then suffers from degraded parallel scaling.  For this version of BGW, we note that the significant intermediate parallel I/O requirements combined with rearrangement of the entries of $\tilde P^q_{g,g'}$ (i.e., memory contention) are the main reasons for the performance dropoff.

Figure~\ref{fig:parallelscaling} also shows the dependence of the OpenAtom performance versus the number of chares used in the $P^q$ computation as described above.  The dependence on the number of chares is modest but non-negligible.  For smaller number of nodes, a more modest number of chares generates the best performance, while for large node counts, enlarged numbers of chares enhance performance.

\begin{center}
\begin{figure*}
\centering
\subcaptionbox{IBM BlueGeneQ:  Mira}{\includegraphics[width=300pt, angle=0]{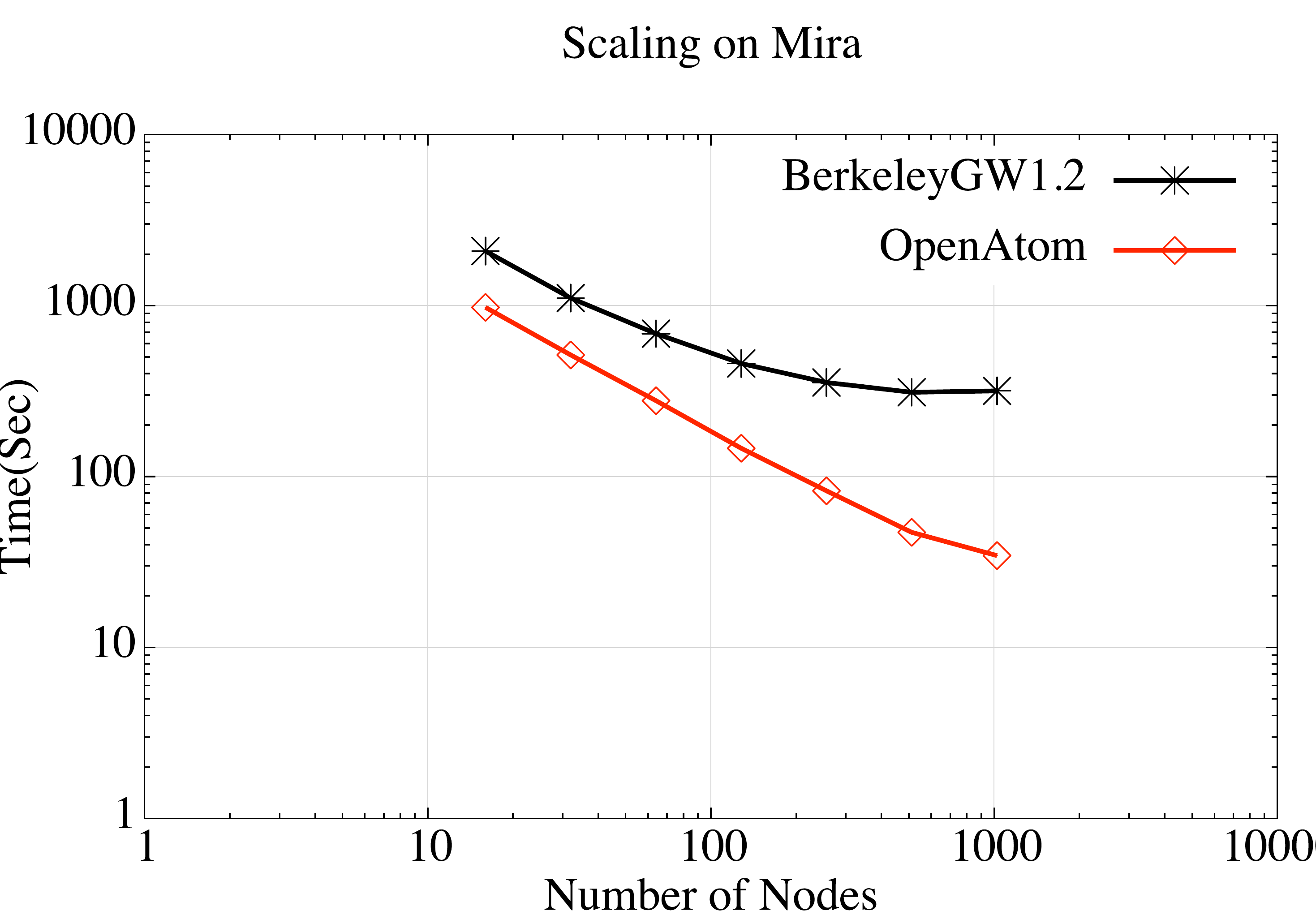}}
\subcaptionbox{Cray XE : Bluewaters}{\includegraphics[width=300pt, angle=0]{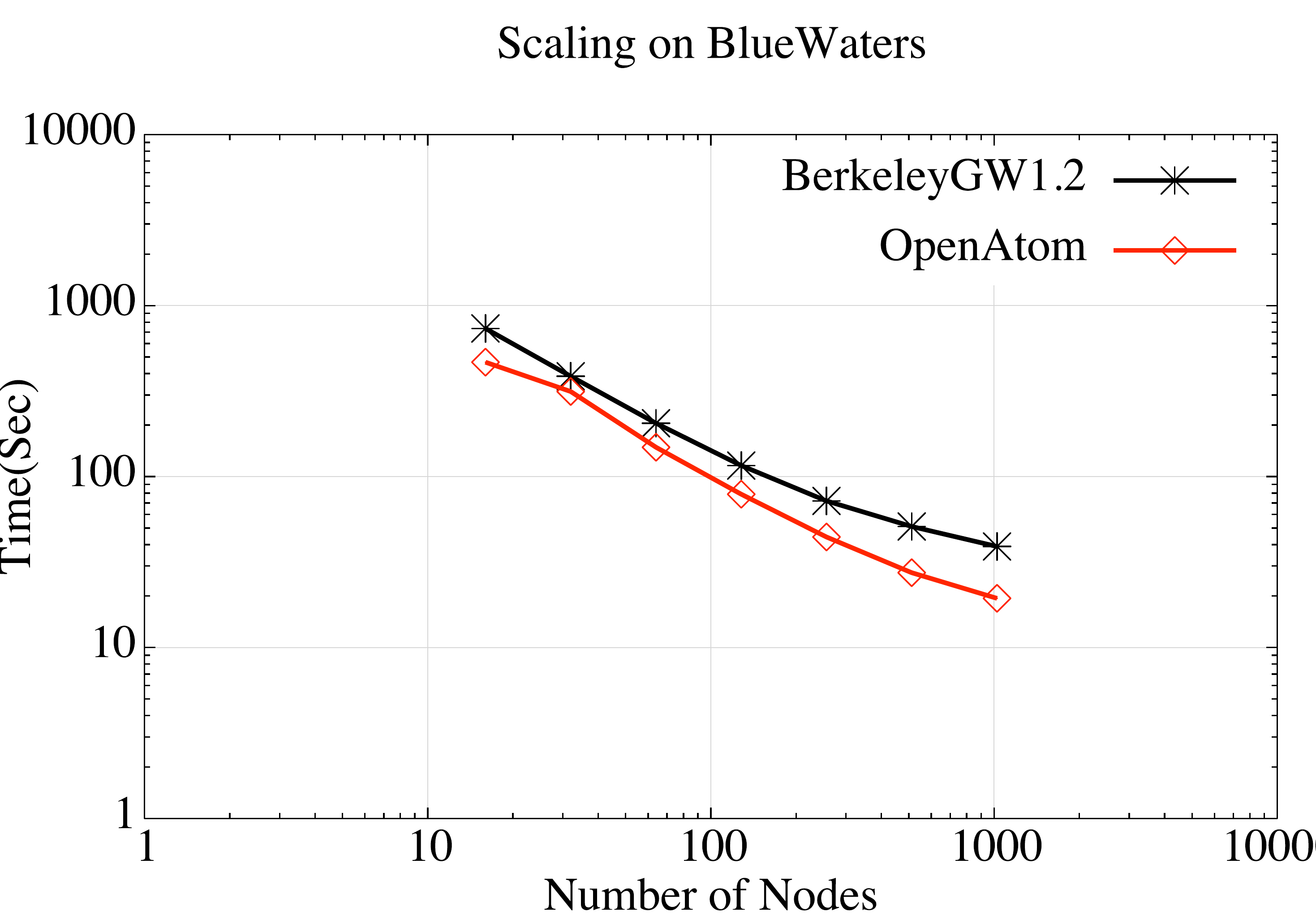}}
\caption{Performance scaling of the polarizability matrix calculation with the BerkeleyGW (BGW) and OpenAtom (OA) software packages on two HPC platforms with 32 threads per node.  The system is bulk Si with a 108 atom unit cell, 12 Ryd plane wave cutoff, and 10 Ryd cutoff for the polarizability. This shows a strong scaling results generating the same convergence level. 
 88$\times$88 chares are used by OA for these results.
}
\label{fig:parallelscaling108}
\end{figure*} 
\end{center}

For the ``medium'' sized system, we analyzed scaling performance with 108 Si atoms for 1 kpoint. The size of the FFT grid for this system is 42$\times$22$\times$22.  The computing time for $P^q$ evaluated at $q=(0.0,0.0,0.001)$ is plotted as a function of number of nodes in Figure ~\ref{fig:parallelscaling108}. This is obtained on IBM BlueGene, Mira with each node consisting of 32 threads. While BGW (in black) stops scaling after 500 nodes, \openatom{} (in red) scales linearly till 1024 nodes. We find that on 1024 nodes, \openatom{} is one order magnitude faster than BGW. We observe that \openatom{} scales similarly on Blue Waters as well. On 1024 nodes, \openatom{} is 3 times faster than BGW on Blue Waters. 

Several optimizations discussed in section~\ref{problemDecomposition} were implemented to improve scaling of \openatom{} on Mira and Blue Waters. During the computation of $P^q$, multiple unoccupied states are broadcast to nodes to perform more $f$ vector computations to increase processor utilization and significantly improved performance by 10$\times$ on large node counts. In addition, computing only those $f$ vector entries required by matrix tiles on each node resulted in further improvement of computation time of up to 3$\times$ on larger node counts.

For the ``large'' sized system we analyzed scaling performance with 432 Si atoms for 1 kpoint. The size of the FFT grid for this system is 42$\times$42$\times$42.  We compare the computing time for $P^q$ evaluated at $q=(0.0,0.0,0.001)$ for BGW and \openatom{} on IBM BlueGene, Mira with each node consisting 16 threads. On 512 and 1024 nodes BGW  takes
13564 and 7048 seconds respectively. \openatom{}  takes almost half of the time; 6900 and 3771 seconds for 512 and 1024 nodes respectively. \openatom{} also shows a strong scaling with 3072 nodes or $\sim$ 50K cores. It is worth mentioning that with 32 threads per node on 1024 nodes \openatom{} improves to 2550 seconds, whereas BGW ran out of memory.    

\section{Conclusions}

The GW method is a powerful tool for the evaluation of quasi-particle properties of complex many-body systems.  However, for large systems, supercomputers and corresponding parallel software and algorithms are required to achieve results in a useful time frame. In order to address these large systems with reduced computational effort, we have presented a complete analysis of the standard formulae, in a variety of representations, to determine through complexity analysis the best expressions for numerical evaluation.  Using the most efficient formulations, we then develop a new massively parallel application on top of the \charmpp{} parallel middleware and demonstrated highly effective performance on a range of systems sizes from large to small.

In more detail, we reanalyze the variant of the GW method that uses plane-wave based DFT for the baseline input, and perform a detailed computational complexity analysis considering both real-space and reciprocal space representations of all method components. The analysis reveals that some terms commonly computed in the $g$-space representation can be expressed in real-space to achieve more efficient GW computations.  In particular, the operation count for the static polarization matrix and the inverse frequency dependent dielectric constant can be greatly reduced, leading to large performance improvements.

Following our earlier work on Car-Parinello DFT 
\cite{vadali_jcc_2004,bohm_ibmjresdev_2008,bhatel_europar_2009}, we implemented GW using the most efficient representation of the standard formulae on top of the \charmpp{} parallel middleware. This approach frees application development from the details of supercomputer platform architecture by using over-decomposition of virtual parallel objects and has been shown to achieve high paralleling scaling on many applications. We have been able to verify our application versus standard software and have shown significant improvements in parallel scaling on a range of systems on the large scale parallel machines Mira and Blue Waters.  Our new open software is freely available at \url{http://charm.cs.illinois.edu/OpenAtom} for use by the community.

In addition to generating massively parallelized GW software, what would be useful to the community is reducing the cost of the GW computations. To this end, we have developed the new $O(N^3)$ scaling $P$ and $\Sigma(x,x',\omega)$ calculation algorithms based on the real-space method~\cite{our_cubic_scaling}. We showed our $O(N^3)$ method significantly reduces the computational loads, which is more than a few orders of magnitude less computation compared to the standard $O(N^4)$ method: the bigger the system, the higher the savings.
Thus, we plan to implement our new $O(N^3)$ method into \openatom{} so that the most time consuming $P$ calculation and the full frequency $\Sigma$ calculation scale as $O(N^3)$. Having an extremely well parallelized software that implements a cubic scaling GW algorithm will have a significant impact to the community.

\section*{Acknowledgement}
This work was supported primarily by the National Science Foundation via the grants NSF ACI-1339804 and 1339715.  Computational resources on Blue Waters were enabled via the grant NSF OAC-1614491.
This research also used resources of the Argonne Leadership Computing Facility at Argonne National Laboratory, which is supported by the Office of Science of the U.S. Department of Energy under contract DE-AC02-06CH11357. G. J. M. was supported by Pimpernel Science Software and Information Technology.
\newpage

%\bibliographystyle{unsrt}
%\bibliography{refs}

\end{document}